\documentclass[useAMS,usegraphicx,usenatbib]{mn2e}
\usepackage{amsbsy,amsmath,amssymb,amstext,longtable,times,xspace}
\usepackage{natbib,ulem,xcolor}

\title[Inactivity of the retrograde-planet host \nuoa]{A photospheric and 
chromospheric activity analysis of the quiescent retrograde-planet host 
\mbox{$\nu$ Octantis~A}}

\author[D.~J.~Ramm \oo]{$\rm D.~J.~Ramm^1$\thanks{E-mail: djr1817@gmail.com}\thanks{Research Fellow}, 
$\rm P.~Robertson^2$, $\rm S.~Reffert^3$, $\rm F.~Gunn^1$, 
$\rm T.~Trifonov^4$, $\rm K.~Pollard^1$, \and
$\rm F.~Cantalloube^4$ \\
$^1$ School of Physical and Chemical Sciences, University of 
Canterbury, Private Bag 4800, Christchurch 8140, New Zealand \\
$^2$ Department of Physics \& Astronomy, The University of California, Irvine, 
Irvine, CA 92697, USA \\
$^3$ Landessternwarte, Zentrum f{\"u}r Astronomie der Universit{\"a}t 
Heidelberg, K{\"o}nigstuhl 12, 69117, Heidelberg, Germany \\
$^4$ Max-Planck-Institut f{\"u}r Astronomie, K{\"o}nigstuhl 17, Heidelberg 
69117, Germany
}

\newcommand*{\nc}{\newcommand*}

\nc{\esm}{\ensuremath}
\nc{\xs}{\xspace}
\nc{\oo}{et~al.\xs}

\nc{\nuo}{\esm{\nu}~Oct\xs}
\nc{\nuoa}{\esm{\nu~{\rm Oct\,A}}\xs}
\nc{\nuob}{\esm{\nu~{\rm Oct\,B}}\xs}
\nc{\nuoab}{\esm{\nu~{\rm Oct\,Ab}}\xs}
\nc{\gdra}{\esm{\gamma~{\rm Dra}}\xs}
\nc{\gcepa}{\esm{\gamma~{\rm Cep\,A}}\xs}
\nc{\gcepab}{\esm{\gamma~{\rm Cep\,Ab}}\xs}

\nc{\se}{\section}
\nc{\su}{\subsection}
\nc{\suu}{\subsubsection}
\nc{\noi}{\\ \noindent}
\nc{\non}{\nonumber}

\nc{\sss}{\rm \scriptscriptstyle}
\nc{\scs}{\rm \scriptstyle}
\nc{\lab}{\label}

\nc{\ml}{\multicolumn}

\nc{\mc}{\mathcal}
\nc{\cs}{\esm{\mc{M}_2}\xs}
\nc{\mjup}{\esm{\mc{M}_{\rm Jup}}\xs}
\nc{\kfour}{\esm{\rm 4k\!\times\!4k\ \xs}}
\nc{\eps}{\esm{\epsilon}\xs}

\nc{\prot}{\esm{P_{_{\rm rot}}}\xs}
\nc{\porb}{\esm{P_{_{\rm orb}}}\xs}
\nc{\prv}{\esm{P_{_{\rm RV}}}\xs}

\nc{\lum}{\esm{\mc L}\xs}
\nc{\rad}{\esm{\mc R}\xs}
\nc{\mass}{\esm{\mc M}\xs}
\nc{\msun}{\esm{~\mc{M}_{\odot}}\xs}
\nc{\lsun}{\esm{~\mc{L}_{\odot}}\xs}
\nc{\rsun}{\esm{~\mc{R}_{\odot}}\xs}
\nc{\mbv}{\esm{M_{\rm V}}\xs}

\nc{\teff}{\esm{T_{\rm eff}}\xs}
\nc{\tmldr}{\esm{T_{_{\rm MLDR}}}\xs}
\nc{\mldr}{\esm{\Delta m_{_{\rm LDR}}}\xs}
\nc{\mhip}{\esm{\Delta m_{_{\rm Hip}}}\xs}
\nc{\sldr}{\esm{\sigma_{_{\rm LDR}}}\xs}
\nc{\ship}{\esm{\sigma_{_{\rm Hip}}}\xs}

\nc{\redchi}{\esm{\chi ^2_{_\nu}}\xs}
\nc{\iod}{\esm{\rm I_2}\xs}

\nc{\dbin}{_{\sss bin}}
\nc{\dpl}{_{\sss pl}}
\nc{\apl}{\esm {a\dpl}\xs}
\nc{\epl}{\esm {e\dpl}\xs}
\nc{\abin}{\esm {a\dbin}\xs}
\nc{\ebin}{\esm {e\dbin}\xs}
\nc{\pbin}{\esm{P\dbin}\xs}
\nc{\ppl}{\esm{P\dpl}\xs}
\nc{\ibin}{\esm{i\dbin}\xs}
\nc{\irot}{\esm{i_{\rm rot}}\xs}
\nc{\ipl}{\esm{i\dpl}\xs}
\nc{\dg}{\esm{\degr}\xs}

\nc{\kms}{\esm{\rm km\,s^{-1}}\xs}
\nc{\ms}{\esm{~\rm m\,s^{-1}}\xs}

\nc{\hc}{HERCULES\xs}
\nc{\hip}{{\it Hipparcos}\xs}

\nc{\ie}{i.e.~}
\nc{\eg}{e.g.~}
\nc{\fn}{\footnote}

\nc{\bc}{\begin{center}}
\nc{\ec}{\end{center}}
\nc{\bte}{\begin{table}}
\nc{\ete}{\end{table}}
\nc{\btr}{\begin{tabular}}
\nc{\etr}{\end{tabular}}
\nc{\bfi}{\begin{figure}}
\nc{\efi}{\end{figure}}
\nc{\beq}{\begin{equation}}
\nc{\eeq}{\end{equation}}

\nc{\eql}{Eq.~\eqref}
\nc{\fl}{Fig.~\ref}
\nc{\tl}{Table~\ref}
\nc{\scl}{\S~\ref}
\nc{\fnl}{Footnote~\ref}

\nc{\stn}{\esm{S/N}\xs}
\nc{\mstn}{\esm{\langle S/N\rangle}\xs}
\nc{\mindex}{\esm{\langle I\rangle}\xs}
\nc{\si}{\esm{\sigma}\xs}

\nc{\hal}{\esm{\rm{H}\,\alpha}\xs}
\nc{\caii}{\esm{\rm Ca\,{\scs II}}\xs}
\nc{\cai}{\esm{\rm Ca\,{\scs I}}\xs}
\nc{\rvcai}{\esm{RV_{\rm abs}}\xs}

\nc{\di}{_{\rm i}}

\nc{\fzero}{\esm{F_{\sss 0}}\xs}
\nc{\fone}{\esm{F_{\sss 1}}\xs}
\nc{\ftwo}{\esm{F_{\sss 2}}\xs}
\nc{\ang}{\,\AA\xs}
\nc{\mnFo}{\esm{\langle F_0 \rangle}\xs}
\nc{\mnFi}{\esm{\langle F\di \rangle}\xs}
\nc{\dmag}{\esm{\Delta m}\xs}
\nc{\snth}{\esm{S_{\rm cut}}\xs}

\begin{document}

\pagerange{\pageref{firstpage}[2793]--\pageref{lastpage}[2806]}
\pubyear{2021}

\maketitle

\begin{abstract}
The single-lined spectroscopic binary $\nu$~Octantis provided evidence of the 
first conjectured circumstellar planet demanding an orbit retrograde to the 
stellar orbits. The planet-like behaviour is now based on 1437 radial 
velocities 
(RVs) acquired from 2001 to 2013. \nuo's semimajor axis is only 2.6~au with the 
candidate planet orbiting \nuoa about midway between. These details seriously 
challenge our understanding of planet formation and our decisive modelling of 
orbit reconfiguration and stability scenarios. However, all non-planetary 
explanations are also inconsistent with numerous qualitative and quantitative 
tests including 
previous spectroscopic studies of bisectors and line-depth ratios, photometry 
from \hip and the more recent space missions TESS and {\it Gaia} (whose increased 
parallax classifies \nuoa closer still to a subgiant, $\sim$~K1\,IV). We 
conducted the first large survey of \nuoa's 
chromosphere: 198 \caii H-line and 1160 \hal indices using spectra 
from a previous RV campaign (2009--2013). We also acquired 135 spectra 
(2018--2020) primarily used for additional line-depth ratios, which are 
extremely sensitive to the photosphere's temperature. We found no significant 
RV-correlated variability. Our line-depth ratios indicate temperature 
variations of only $\pm4$~K, as achieved previously. Our atypical \caii 
analysis models the indices in terms of \stn and includes covariance 
significantly in their errors. The \hal indices have a quasi-periodic 
variability which we demonstrate is due to telluric lines. Our new evidence 
provides further multiple arguments realistically only in favor of the planet.
\end{abstract}


\se{Introduction}
\lab{sect:intro}
A decade or so after the first exoplanets were described 
(Wolszczan \& Frail 1992; Mayor \& Queloz 1995), radial velocity (RV) evidence 
for the first retrograde planet appeared from a very unexpected source, the 
compact single-lined spectroscopic binary $\nu$ Octantis (Ramm 2004). However, 
these RVs shared the same initial fate as those from \gcepa which almost hosted 
the first acknowledged RV-discovered planet (Campbell, Walker \& Young 1988; 
Walker \oo 1992). \gcepab was eventually confirmed 15~yr later (Hatzes \oo 
2003), and was then the shortest period binary with a planet 
($\pbin\sim 57$~yr) -- about 20$\times$ longer than that for \nuo. Both 
initial series of RVs led their discoverers to describe both planet and stellar 
rotation-related scenarios, the latter then being suspected as more likely 
causes. These two binaries have many other parallel details in their exoplanet 
histories and host star characteristics.\fn{For example, and very 
significantly, both hosts were originally classified as giants, rare for early 
exoplanet claims -- and both K0\,III -- which made a stellar origin for the RV 
signals more tenable. They also share a trivial near-polar declination 
detail: $\lvert\delta\rvert = 77\dg$.}

Host stars may provide evidence that a candidate exoplanet orbit is retrograde 
relative to the star's rotation (which we label type-1), and if 
in a binary system, relative to the two stellar 
orbits (our type-2). The first acknowledged retrograde planet was HAT-P-7b, 
promptly identified using the McLaughlin-Rossiter effect (our type-1; Winn \oo 
2009; Narita \oo 2009). This had been preceded in the same year by the first 
paper describing the conjectured planet \nuoab, (Ramm \oo 2009; henceforth 
R09), which has proven to be much more challenging for establishing its reality 
beyond reasonable doubt. This is almost entirely due to the unprecedented 
geometry of the conjectured system, whose stars are separated by only 2--3~au 
with the circumstellar \ie S-type planet about midway between (see \eg Gong \& 
Ji 2018; Bonavita \& Desidera 2020; Quarles \oo 2020). 
Some details for \nuo (HD\,205478, HIP\,107089) and its conjectured planet 
(based on the persistent RV cycle with a period $\prv\sim 415$~d) are listed in 
\tl{stellar} and \tl{orbital}. The parallax was recently updated from {\it Gaia} 
observations, and critically for the planet claim, increased by about 10 per 
cent (Gaia Collaboration 2018; Kervella \oo 2019 - whose work specifically 
includes nearby stars with stellar and substellar companions).\fn{Thus \nuoa is 
less luminous and should be classified closer still to 
a subgiant, $\sim$~K1\,IV, as now is also \gcepa (Hatzes \oo 2003; Fuhrmann 
2004). \scl{sec:gaia} will discuss this important revision.}

The system's geometry makes a prograde planet impossible (R09; Eberle \& Cuntz 
2010). An S2 (\ie S-type, our type-2 retrograde) orbit has a significantly 
wider stability zone than its prograde equivalent (see \eg Jefferys 1974; 
Wiegert \& Holman 1997; Morais \& Giuppone 2012).\fn{With zero evidence for 
retrograde status we label the orbit type-0 \ie prograde, and assumed if not 
stated. If the host provides evidence of both type-1 and 2 we would label it 
type-3 (1+2). The scheme is applicable to single star hosts \ie type 0 or 1, 
and both S- (circumstellar) and P-type (circumbinary) planets (Dvorak 
1986), \ie type 0, 1, 2 or 3.} This opportunity for \nuoab's reality was first 
investigated by Eberle \& Cuntz (2010), and subsequently by Quarles, Cuntz \& 
Musielak (2012), Go{\'z}dziewski \oo (2013) and Ramm \oo (2016; henceforth 
R16), all of which found retrograde solutions with merit but without being 
conclusive. Meanwhile, however, multiple other tests demonstrate that all 
other explanations (\eg measurement artefacts, stellar variability, \nuo being 
a triple-star system) are substantially less credible than the planet (R09; 
Ramm 2015 - henceforth R15; and R16). If the planet is an illusion, \nuoa so 
far provides no evidence of its causative role other than 
the persistent precise cycle of so far 1437 RVs over 12.5~yr, and thus would 
have the alternative distinction of seriously confronting our understanding of 
stellar variability.\fn{The RVs were 
obtained using two CCDs and different calibration techniques (\iod-cell and 
thorium-argon spectra) and therefore also different reduction methods. No other 
star's study using our instruments and methods provide any similar planet-like 
RVs. The primary star's argument of periastron, $\omega_1=75\dg$, has not 
changed significantly in 95~yr so any claim that a hierarchical triple-star 
system is creating the planet-like RVs is also unfounded (see R09 and R16 for 
details).\lab{tenuous}}

\bte
\bc
\btr{lcc}
\hline
  Parameter       &  $\nu$ Octantis A             & Reference \\ 
\hline
Spectral type$^{\rm a}$  &   K1IIIb-IV          & (1) \\
$V$ (mag)         &  $+3.738\pm0.005$             & (2) \\
parallax (mas)    &  $51.546\pm0.705$             &  (3) \\
$\mbv$ (mag)      &  $+2.30\pm0.16$               & our calculation \\ 
$BC_{\rm V}$ (mag) &  $-0.39\pm0.05$                & (2) \\
$H_{\rm p}$ (\hip mag)  &  $3.8981\pm0.0004$        & (4) \\ 
Mass (\msun)$^{\rm a}$  &  $1.61\pm0.15$, $1.4\pm0.3$   &  (2), (5) \\ 
Radius (\rsun)$^{\rm a}$  &  $5.81\pm0.12$, $6.0\pm0.3$   &  (2), (5) \\ 
\teff (K)$^{\rm a}$   &  $4860\pm40$, $4815\pm90$     &  (2), (5) \\ 
$T_{_{\rm LDR}} \rm (K)^{\rm b}$  & $4811\pm45$, $4810\pm45$  &  (1), (6) \\
Luminosity (\lsun)$^{\rm a}$ &      $\sim17$        &  (5)  \\
$v\sin i$ (\kms)  &      $2\pm0.5$                  & (2) \\
\prot (d)$^{\rm a}$ &     $\simeq140\pm35$          &   (6)  \\
Age (Gyr)         &  $\sim2.5$--3                 &   (5) \\ 
Observations (RVs)$^{\rm c}$ & 1180 \iod-cell, 257 CCF       & (1), (5) \\
time span of RVs  & $2001.43-2013.96$             & (1), (5) \\ 
\hline							                     
\etr
\caption{Stellar parameters and RV observations for \nuoa (all errors are 
$\pm1\si$, except $H_{\rm p}$ which is the standard error on the median). 
1:~Ramm \oo (2016), 2:~Fuhrmann \& Chini (2012), 3:~Kervella \oo (2019): this 
parallax is a Bayesian ZP-corrected value from ESA's {\it Gaia} mission (GRD2: 
Gaia Collaboration 2018), 4:~ESA (1997), 5:~Ramm \oo (2009), 6:~Ramm (2015). 
$^{\rm a}$ Reviewed in \scl{sec:gaia}. $^{\rm b}$ LDR: line-depth ratio analysis. 
$^{\rm c}$ See \fnl{tenuous}.}
\lab{stellar}
\ec
\ete

\bte
\bc
\btr{lcc}
\hline
Parameter         & Binary               &  planet, Ab          \\   
\hline 							                     
Companion Mass    &  $0.5-0.6$ \msun       &  $2.1-2.6$ \mjup   \\
Object class      & $\rm K7-M0V$, or WD        &  Jovian        \\
$K$               &    7.05 \kms         &  $40-45$\ms     \\ 
$a$  (au)         &    $2.5-2.6$               &  $1.2-1.3$     \\  
$\porb$       &  $\rm 1050.1~d\simeq 2.9~yr$ &  $415-418$~d     \\ 
$e$               &  0.24                &  $0.09-0.13$         \\ 
\hline
\etr
\caption{Basic orbital parameters for \nuo and the conjectured S2-type 
retrograde planet, summarised from R09, Go{\'z}dziewski \oo (2013) and R16. The 
ratio of the orbital periods, \porb, is close to 5:2. Quarles \oo (2012) and 
Go{\'z}dziewski \oo (2013) both favour a coplanar planetary orbit which R16 
could not confirm.}
\lab{orbital}
\ec
\ete

Go{\'z}dziewski \oo identify a small number of nearby mean motion resonances 
for coplanar S2 orbits in \nuo including one at the period ratio 5:2, this 
having been qualitatively suspected in R09. The unprecedented strong 
interactions suggest resonance is likely to be a contributing factor for 
stability, as in other multi-body systems (\eg Campanella 2011; Robertson \oo 
2012; Horner \oo 2019; Stock \oo 2020). \nuoab would never 
be confined to a typical narrow orbital path but would move endlessly 
throughout a rather wide zone as Eberle \& Cuntz (2010) first illustrated, 
leading to significant challenges for competing models trying to characterize 
different orbital geometries (Panichi \oo 2017).

Many S-type planets have now been reported, though demands for any following an 
S2 orbit remains rare since the binary must be quite compact 
to lead to this conclusion. Scenarios that may create such orbits include 
star-hopping (Kratter \& Perets 2012), perhaps facilitated by a scattering 
event, or significant system or orbital changes involving stellar winds, 
accretion/debris discs or evolution to a white dwarf (WD) companion (see \eg 
Tutukov \& Ferorova 2012). The WD scenario may have support for \nuo from 
spectrophotometric observations nearly 50~yr ago (Arkharov, Hagen-Thorn \& 
Ruban 2005). They reported unusually large variability ($S_{\nu}=0.14$~mag) in 
the near ultraviolet (345~nm) in the 1970s. Of the 172 `normal' and 176 
`suspect variable' stars they catalogued, only two others had larger $S_{\nu}$, 
both in the latter group. This may be a clue to the nature of \nuob as it 
seems unlikely \nuoa is the source.

The first close-planet candidate for a WD host was recently reported 
(WD1856+534; Vanderburg \oo 2020), discovered using NASA's Transiting Exoplanet 
Survey Satellite (TESS; Ricker \oo 2015), another space mission that should 
help 
unravel the \nuo mystery. WD1856\,b's orbit ($\ppl\sim1.4$~d), is presumed to 
be due to reconfiguration and migration processes (see \eg Lin, Bodenheimer \& 
Richardson 1996; Rasio \& Ford 1996), which must have relevance for any 
orbitally retrograde planet. Together with mass loss/transfer processes 
inherent in WD evolution, these scenarios may further complicate attempts to 
understand \nuoab but also provide some advantage since orbit reconfiguration 
(of the planet or the binary) would not require interaction from 
external or other internal masses, as anticipated with an unevolved secondary 
star.

Two other recent claims are also relevant here. Firstly, another S2-type 
planet, HD\,59686\,Ab, has been conjectured in this somewhat wider but rather 
eccentric binary (Ortiz \oo 2016; Trifonov \oo 2018; K2\,III, 
$\abin\sim13.6$~au, $\ebin\sim0.7$). Interestingly, HD\,59686\,B is also 
suspected of being a WD, leading Ortiz \oo to investigate one possibility that 
the planet formed from a second generation protoplanetary disc. It is tempting 
to speculate that \nuo and HD\,59686 may be two early examples of 
relatively compact binaries with WD secondaries that created S2-type planets. 
Secondly, the first planet in a binary more compact than \nuo was recently 
reported (HD\,42936, $\rm K0V+L$ dwarf; $\abin\sim1.2$~au; Barnes \oo 2020). 
HD\,42936-Ab is close to its host ($\sim 0.066$~au) so both it and WD1856\,b 
are assumed to have reconfigured orbits, but with no evidence demanding either 
be retrograde.

Understandable suspicion also persists that \nuoa's RV signal may be 
due to undetected surface activity. Hatzes \oo (2018) repeated this 
concern after \gdra's anticipated single-planet claim disappeared after more RV 
data was acquired. Oscillatory convective modes (Saio \oo 2015) were instead 
promoted as the variable RVs more likely origin, just as Reichert \oo (2019) 
speculate for Aldebaran -- instead of its conjectured planet (Hatzes \oo 2015). 
But unlike \nuoa which is a relatively unevolved low-luminosity star (and 
so far has a persistent periodic RV signal), \gdra ($\sim 500$\lsun, 50\rsun), 
and Aldebaran ($\sim 400$\lsun, 40\rsun) typify the theoretical high-luminosity 
expectations of a star having these convective modes. Both are classified 
K5\,III and are reminders of the potential complications such evolved stars, 
both real and mistaken (\eg \nuoa and \gcepa) can present. 
Other radial and non-radial oscillations 
in stars similar to \nuoa instead have a complex power spectrum that complicate 
asteroseismology studies (see \eg Dupret \oo 2009). This is not at all 
consistent with the planet-like RV signal \nuoa has provided to date. Hence, if 
\nuoa is deceiving us, a more reasonable concern is that stars more similar to 
it are more likely to be creating well-crafted illusions, such as \gcepa. 
This possibility would create significant 
complications for other exoplanet claims now and in the future, and perhaps not 
restricted to similar subgiants.

This background has motivated the photospheric and chromospheric studies 
reported here. Photospheric line variations are critical for RVs but also for 
the extremely temperature-sensitive method of line-depth ratios (LDRs; \eg 
Hatzes, Cochran \& Bakker 1998; Gray \& Brown 2001; Kovtyukh \oo 2003; R15; 
R16). R15 and R16 demonstrate temperature constancy for \nuoa to as little as 
$\pm 4$~K over 12~yr, with high consistency with the very low 
photometric variability recorded by \hip (see \tl{stellar}). This latest LDR 
series is also relevant as, more or less concurrently, observations have also 
been obtained by NASA's TESS mission which significantly support \nuoa's 
photometric stability, as well as RVs from ESO's HARPS spectrograph (Trifonov 
\oo, in preparation).

Chromosphere activity of \nuoa has so far been limited to three eye estimates 
of the \caii K line from each of two spectra (Warner 1969), a single spectrum 
of \caii H \& K (R09) and H-line indices from 17 spectra (R16). All 
three papers concluded \nuoa had minimal \caii activity. Additionally, the 
Mg~$h+k$ chromospheric emission-line study by P\'{e}rez Mart\'{i}nez, 
Schr\"{o}der \& Cuntz (2011) found \nuoa's activity to be very near their 
sample's empirically-derived basal flux of 177 cool giants. Here we report on 
two much larger series: 198 line indices from the \caii H line and 1160 \hal 
indices.

Our work allows \nuoa to be characterized in further detail, providing new 
evidence for the planet since significant activity is again refuted. 
Even if the planet is eventually proven to be an illusion, which seems to 
be increasingly unlikely, our work adds to the fundamental knowledge of the 
star at the centre of this persistently challenging system.

\se{Instrumentation and observations}
\lab{sect:obsns}
% see /home/djr83/activitiy_EWs2020paper/complete_1437rvs_all_columns.tbl
The data analysed here were acquired in two series, one about a decade ago and 
the other over the past two years. Both series were obtained at the University 
of Canterbury Mount John Observatory using the 1-m McLellan telescope and \hc, 
a fibre-fed, thermally insulated, vacuum-housed spectrograph (Hearnshaw \oo 
2002). The CCD was a \kfour detector cooled to $\sim -95\dg$\,C which recorded 
the \nuoa spectra with a resolving power $R\sim70\,000$. Further instrument 
details and our reduction methods for obtaining flat-fielded normalized spectra 
can be found in Ramm (2004), where the first tentative mention of the 
conjectured planet is made, and R16. Our reduction software (HRSP; see Skuljan 
2004) automatically provides both vacuum and air wavelengths. To make other 
vacuum-to-air conversions we used the formulae of Morton (2000).

The older series comprises nearly 1200 spectra obtained from 2009.96--2013.96, 
\ie four years, and allowed our chromosphere studies of the \caii H line and 
\hal. The spectra were almost all obtained using an iodine cell that 
reconfirmed the precise-RV planet-like signal (see R16). For our \hal study, 
1160 spectra were included in the final analysis, made 
possible by the higher \stn at these longer wavelengths. The initial sample 
size for our \caii analysis was limited to 700 spectra as the \caii lines were 
not included in the first year or so of observations. More details of these 
data sets will follow in the relevant sections.

A smaller set of 135 spectra were acquired more recently, from 2018.34 until 
2020.42 \ie over about 25 months. As the \iod cell was not used for these 
observations, we could not obtain further high-precision RVs but instead could 
use these spectra to obtain a new series primarily of line-depth ratios. The 
last spectrum obtained (2020.42; JD245\,9002) has the highest \stn of our 
entire series and was acquired to also extend that range of our \caii spectra. 
The time span of all spectra considered here is from the first LDR spectrum, 
2001.85, until this last observation \ie $\sim19$~yr.

\se{Photosphere line-depth ratios}
We first present the results from our line-depth-ratio analysis of the new 
spectra (2018--2020) and compare them to the previously published LDR results 
(2001--20013 ; R15 and R16) as well as again, the \hip photometry (1990--1993). 
The LDRs demonstrate the continued thermal stability of \nuoa's photosphere and 
so are a useful prelude to our chromosphere studies.

\su{Old and new LDRs and corresponding temperatures}
\lab{sect:ldrs}
We used the same methods and pipelines employed in R15 and R16: specifically, 
for each of the 135 new \nuoa spectra we constructed 22 LDRs and their errors 
from ten spectral lines in the wavelength region 6232--6257~\AA. The principal 
equations for these calculations are given in R15.

One significant advantage of LDRs is that they can be used to derive a star's 
temperature -- which brings additional benefits -- using a series of 
calibration stars whose temperatures allow precise 
interpolation (see R15; their fig.~1). Further increases in accuracy and 
precision are achieved by modelling the influences of temperature and 
luminosity (\ie evolution) using linear regression (Press 
\oo 1994; algorithm {\it fitexy}) leading to a modified LDR labelled MLDR (see 
\eg Gray \& Brown 2001; Catalano \oo 2002; R15). R15 reported an accuracy of 
$45\pm25$~K for recovering the 20 calibration stars' temperatures but a 
precision as small as $\pm4$~K for their 215 \nuoa LDRs, in excellent agreement 
with Gray \& Brown (2001): their temperature precision from 92 giant stars was 
3.9~K. Neglecting these influences can yield significantly poorer temperature 
statistics as Gray \& Brown discuss in the context of other studies.

R15 summarizes the strong relationship between 
the $T-LDR$ regression line's slope $s_{\rm reg}$ and the ratio's error 
$\eps_{\rm r}$ (see their sec.~4.1). The slope is an indicator of sensitivity of 
each ratio to the calibration stars' temperatures. It therefore provides the 
error on each ratio's temperature estimate \ie 
$\eps_{_{T_{\rm r}}}=s_{\rm reg}\times\eps_{_{\rm r}}$. \fl{mldrerrors}\,(a) shows the 
error on each LDR eventually rises exponentially as the spectrum's \stn drops, 
as anticipated. We estimated the \stn solely from the photon noise so that 
$\stn\sim\sqrt{F_{\rm c}}$ where $F_{\rm c}$ is the continuum flux at the spectral 
line, and calculated each ratio's \stn from the average of the two lines. Our 
two examples show typical 
similarities and differences (the relationship illustrated here for the first 
time), and how $\eps_{_{T_{\rm r}}}$ can vary between ratios even though all the 
lines are within about 25\ang ($\sim 6232-6257$\ang).

The 22 \tmldr values next allow the mean and standard deviation, \si, for each 
spectrum to be derived, weighted by $w_{\rm r}=1/\eps_{_{T_{\rm r}}}^2$. Our average 
temperatures have no adjustment for any zero-point differences of the \tmldr 
values for the 22 ratios. The mean values, $\langle\tmldr\rangle$, and their 
standard errors ($S.E.=\si/\sqrt{22}$) are illustrated in \fl{mldrerrors}\,(b) 
and (c) respectively. The final standard deviations have an average 
$30\pm3$~K which is consistent with the temperature accuracy stated above.

\bfi
\bc
{\scalebox{0.4}{\includegraphics{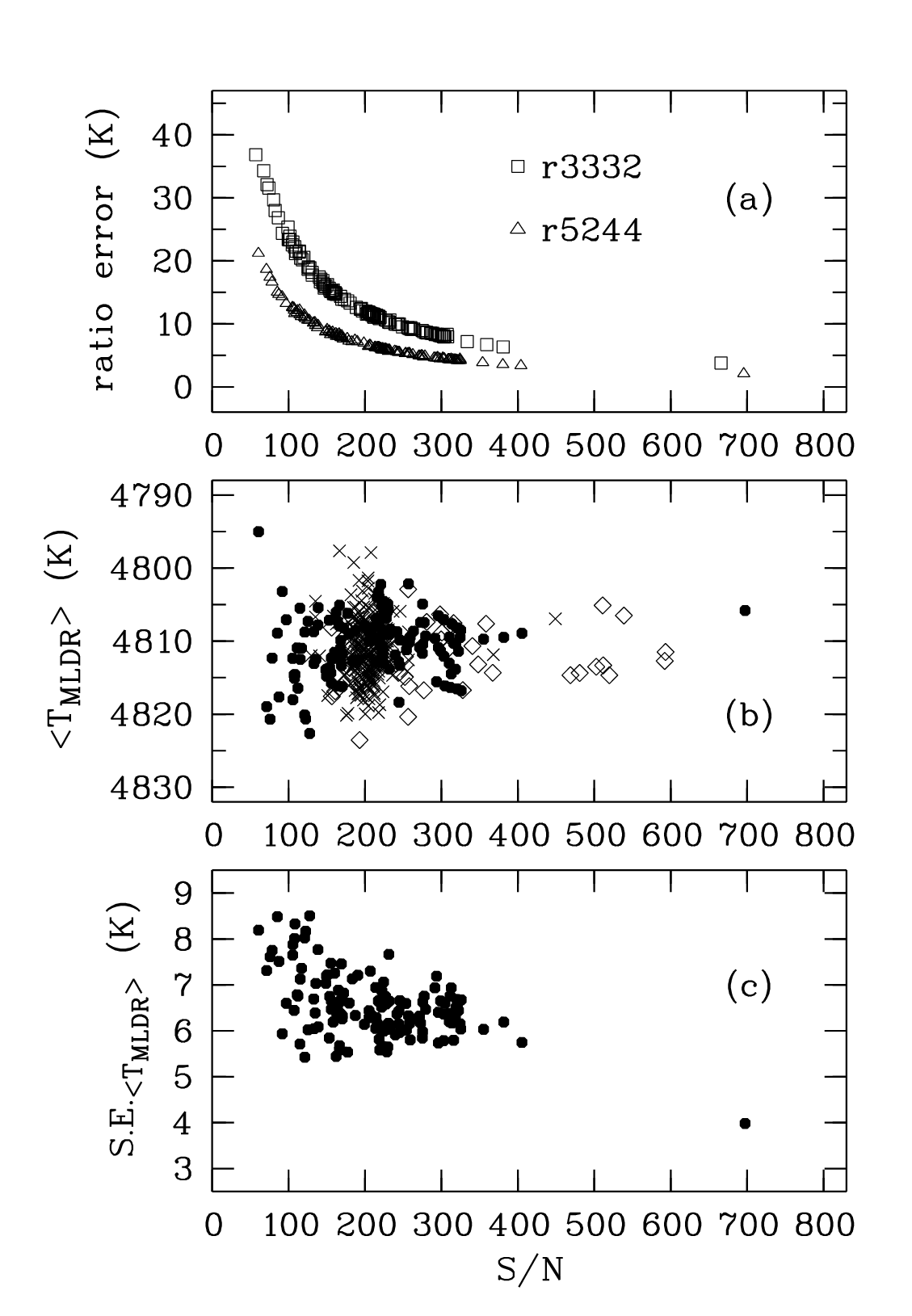}}}
\caption{Temperatures (\tmldr) and errors from modified line-depth ratios 
in terms of \stn. (a) Temperature-calibrated errors, $\eps_{_{T_{\rm r}}}$, for two 
ratios; \eg `3332' labels the line pair 6233\ang and 6232\ang. (b) Averaged 
\tmldr values for 398 spectra: `$\times$' 217 LDRs (2001--2007), 
`$\Diamond$' 46 LDRs (2011--2013), `$\bullet$' 135 LDRs (2018--2020). (c) 
Standard errors on $\langle\tmldr\rangle$. (a) and (c) include only the 135 new 
spectra.}
\lab{mldrerrors}
\ec
\efi

The total set of 398 \tmldr values comprises 217 spectra from the first 
observing series (2001.85--2007.16, 84 nights; R15 and R16), 46 spectra 
acquired during the \iod-RV campaign but without the 
\iod cell (2011.22--2013.74, 20 nights; R16), and our data (26 nights). 
The distribution of all of these LDRs in terms of \stn is provided in 
\fl{mldrerrors}\,(b). This shows that \tmldr 
becomes less scattered as \stn increases, particularly for $\stn\gtrsim250$, 
though this may just be an artefact of the smaller sample size. It can 
also be seen that all of the spectra with $\stn\lesssim120$ were obtained in 
our latest series.

\su{Final temperature statistics and periodogram}
The more recent 135 spectra give a weighted mean temperature 
$T_{\rm \nu Oct}=4810\pm4$~K, with no significant difference from the values 
reported in R15 and R16. Our errors yield $\redchi\sim 0.7$ which indicates 
their average ($\sim6$~K) is somewhat more than what would be consistent with 
our tiny temperature variations. The three series combined have a weighted mean 
$T_{_{\rm \nu Oct}}=4811\pm4$~K. If the temperature-recovery accuracy given above 
($45\pm25$~K) is added in quadrature to this, the final temperature estimate 
for the 2018--2020 spectra is $4810\pm45$~K, equalling the other two LDR 
temperatures in \tl{stellar} -- Ref.\,(1) and (6).

\fl{LDRperiodogram} illustrates the distributions of the mean \tmldr values for 
the 130 epochs of the LDR spectra acquired over 18.6~yr. The means are weighted 
by the standard errors $w=1/\eps_{_{T_{\star}}}^2$ where we abbreviate \tmldr with 
$T_{\star}$. Our results display an extraordinary degree of relative accuracy 
between the three \tmldr series -- the offset of the three mean temperatures is 
remarkably tiny at about 1~K (see $T_{\rm \nu Oct}$ values in 
\fl{LDRperiodogram}). So whilst the absolute accuracy of LDRs may be less 
impressive, there is extraordinary relative accuracy even though their 
sensitivity may be imagined to make them vulnerable to variability from, for 
instance, measurement artefacts including subtle possibilities such as stray 
light within the spectrograph (which is unexpected given the design of \hc and 
that our line-depths create ratios). Clearly our methods provide practically 
identical MLDR distributions of outstanding precision over nearly 20~yr, 
despite two CCDs being used and many other instrumental and environmental 
parameters varying as well. Thus we can be quite confident our LDRs are highly 
unlikely to be significantly compromised by any of these non-stellar variables.

We searched our data for periodicity using 
the generalized Lomb-Scargle (GLS) periodogram as derived by Zechmeister \& 
K{\"u}rster (2009). The GLS is ideal for time series with uneven temporal 
sampling and nonuniform measurement errors as reported here. Our periodograms 
are normalized assuming the noise is Gaussian (Zechmeister \& K{\"u}rster; 
their eq.~22).  In \fl{LDRperiodogram}, we show the result for our 
epoch-averaged MLDR temperatures, along with approximate analytical
$P$-values estimated according to Sturrock \& Scargle (2010). We find no
evidence for statistically significant periodicity at any frequency.

\bfi
\bc
\rotatebox{-90}{\scalebox{0.35}{\includegraphics{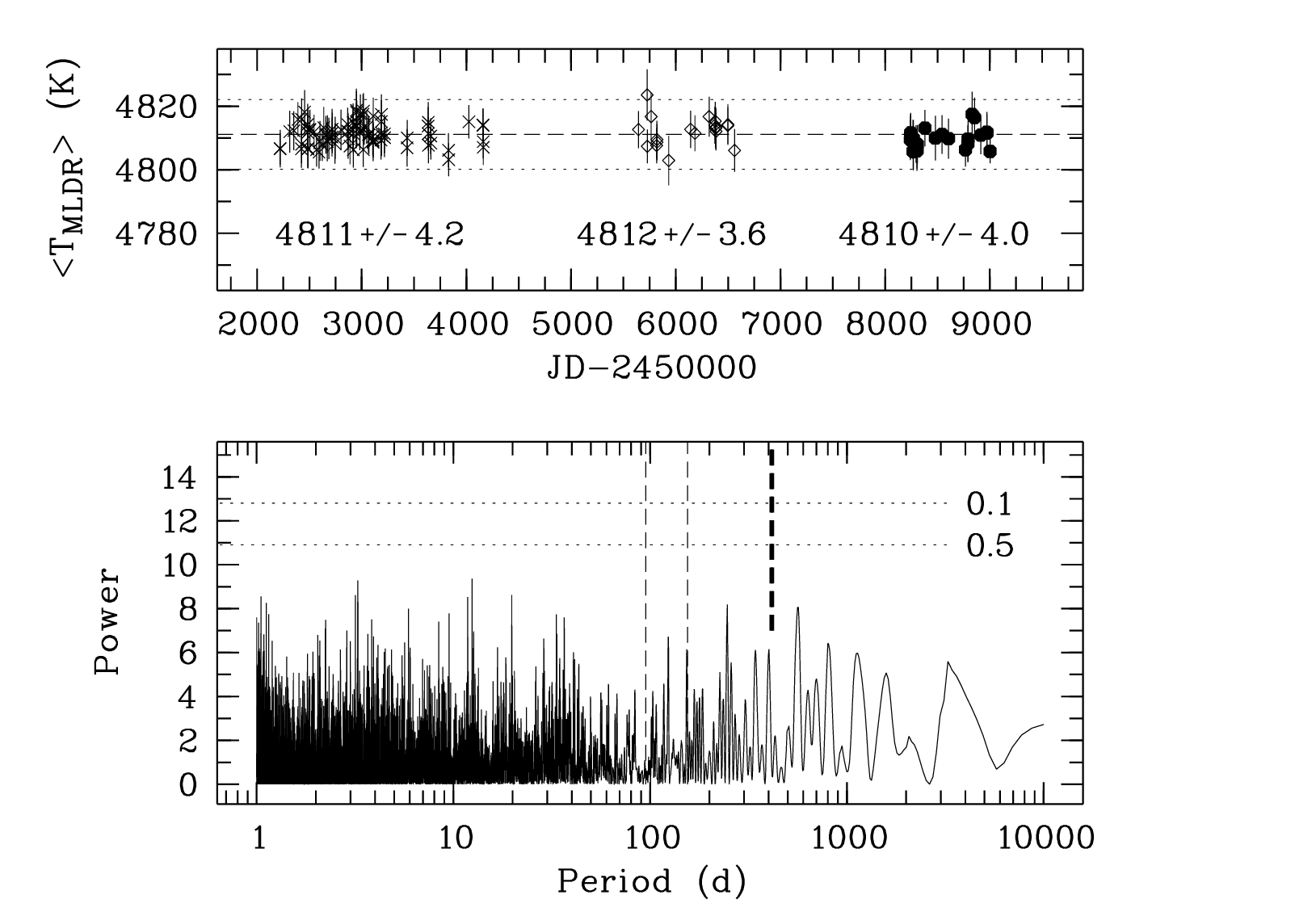}}}
\caption{The 130 nightly means of the \tmldr values from 398 LDR temperatures. 
Top: The distributions of the three LDR series defined in \fl{mldrerrors} 
together with their $\pm1\sigma$ error bars. The mean temperatures 
$T_{\rm \nu Oct}$ and their standard deviations are given below each series. 
Bottom: The Generalized Lomb-Scargle periodogram and corresponding FAP levels. 
The pair of vertical dashed lines identify our revised $\pm 1\sigma$ range of 
\nuoa's predicted rotation period (see \scl{sec:gaia}). The shorter bold 
line marks the planet-like RV period ($\prv \sim415$~d).}
\lab{LDRperiodogram}
\ec
\efi

\su{Photometry from \hip and TESS}
We next make a brief digression to describe photometric satellite data 
which we will then compare to our LDR evidence. Since it has been estimated 
that \nuob is at least 6~mag fainter than \nuoa (R09, and see 
\tl{orbital}), the LDRs and photometry are assumed to only record \nuoa.

Two satellites observed \nuo, acquiring data whose end dates are separated by 
30.6~yr. The first, \hip, achieved the longest baseline ($\sim1176$~d, 
1989.45--1993.17; ESA 1997) and identified \nuoa as one of its more stable 
targets (see \tl{stellar}). More recently, NASA's TESS mission (Ricker \oo 
2015) observed \nuo twice in the 2-minute short-cadence mode. It was 
covered in Sectors 13 (2019 Jun 19 -- Jul 18) and 27 (2020 Jul 5 -- Jul 30), 
\ie baselines of about 28 and 24 days. We retrieved the Pre-search Data 
Conditioning Simple Aperture Photometry (PDCSAP) using the \textsc{Lightkurve} 
software package (Lightkurve Collaboration, 2018). Approximately, this 
photometry has midtimes separated by 395~d and span 408~d. Therefore the 
two TESS records sample slightly different phases of the RV cycle (assuming 
$\prv \sim 415$~d is still relevant), and together they include about 13 
per cent of it. These high-precision records are also consistent with this 
bright spectroscopic binary being exceptionally quiet, at least during these 
short baselines and within TESS's spectral response capabilities (600--1000~nm 
bandpass; Ricker \oo 2015): the RMS scatter of each PDCSAP dataset and both 
combined is just 0.02 per cent.

\bfi
\bc
\rotatebox{-90}{\scalebox{0.3}{\includegraphics{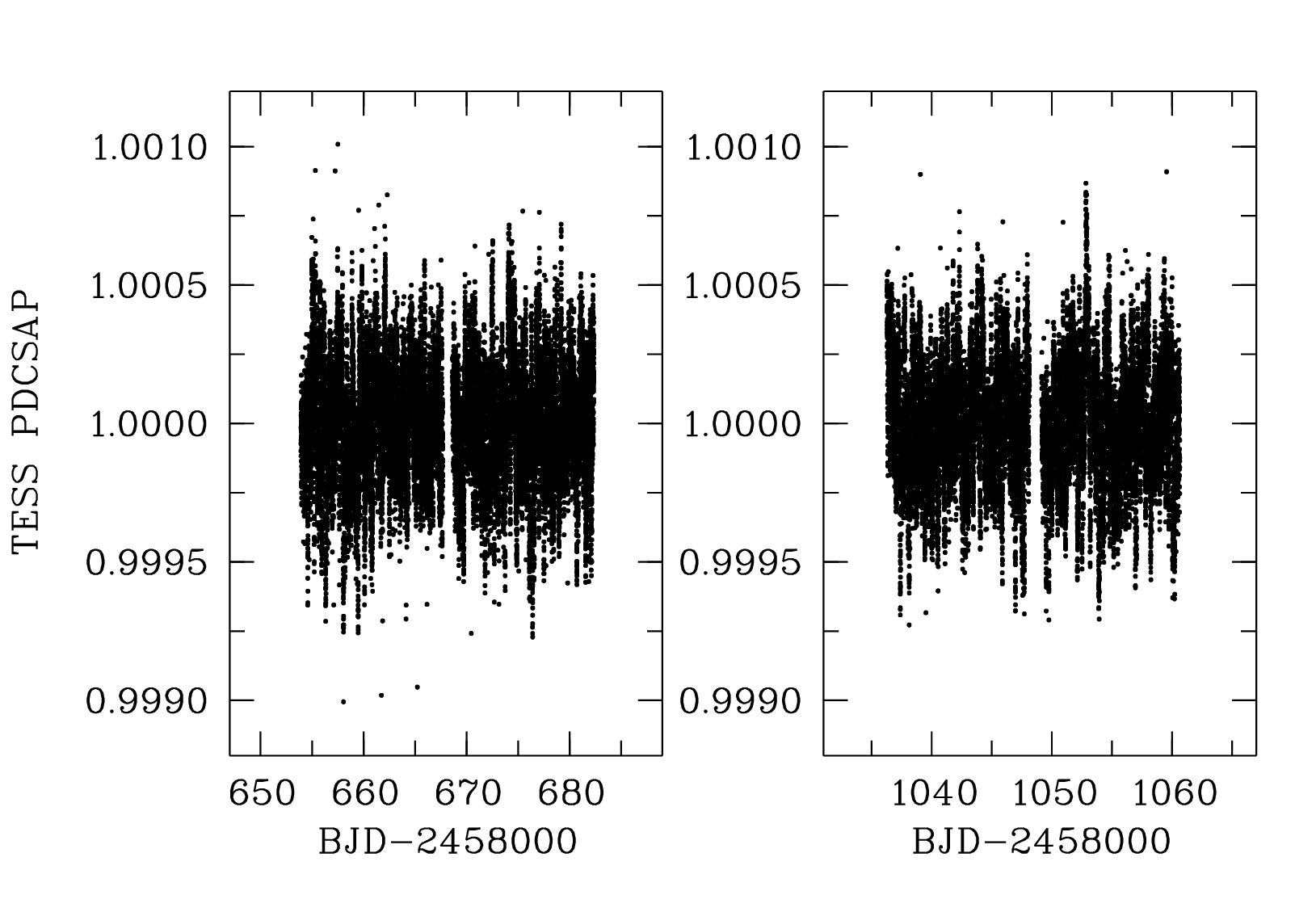}}}
\caption{Two time series for \nuo acquired by NASA's TESS mission during 
mid-2019 ($N=19\,579$) and mid-2020 ($N=16\,780$). The RMS of each data set is 
0.02\%. PDCSAP: Pre-search Data Conditioning Simple Aperture Photometry.}
\lab{tess}
\ec
\efi

\su{Magnitude differences \mldr from \tmldr}
\lab{sec:magdiff}
The temperature calibration enables one more significant benefit: the \tmldr 
values can be converted to a magnitude difference \mldr 
using the Stefan-Boltzmann law ($L\sim R^2T^4$). The \hip and TESS photometry 
strongly support a claim that $\Delta L\sim 0$, and the LDR results just as 
strongly record $\Delta T\sim 0$. Therefore, though \nuoa no doubt has some 
surface variability, as even less evolved stars do, we assume any 
radial changes are insignificant in our next calculation, a repeat of what was 
reported in R15 and R16. \eql{stefanb} provides the conversion to \mldr with 
that assumption (\ie $\Delta R=0$):

\beq
\mldr = -10\log_{10}\left(\frac{T_{\star}}{T_{\nu}}\right)
\pm \frac{10}{\ln(10)}\sqrt{\left(\frac{\si_{_{T_{\star}}}}{T_{\star}}\right)^2 + \left(\frac{\si_{_{T_{\nu}}}}{T_{\nu}}\right)^2}\ ,
\lab{stefanb}
\eeq
\noi
where we abbreviate $T_{\rm \nu Oct}$ (\ie $4810\pm4$~K) with $T_{\nu}$. Note that 
\mldr is extremely sensitive to the temperature: varying 
$T_{\star}$ by only one degree changes \mldr by about one millimagnitude.

In \fl{hipptmldrs} we compare these \mldr estimates to the magnitude 
differences for the longer baseline photometry of \hip, whose \mhip values 
relate to their mean. A striking similarity is evident between the four \dmag 
distributions, all the more so as there 
are no zero-point offsets applied to any \tmldr data. The latest MLDR series 
extends the time span of all \dmag values to about 30.5~yr. This graph also 
demonstrates that the more recent lower-\stn data do not significantly 
compromise the precision of these \dmag values relative to the earlier data 
sets.

Finally we note the small difference between the standard deviations of \mhip 
and \mldr, where $\ship\sim 4.1$ is greater than $\sldr\sim 3.7$~mmag. Of the 
various possible reasons one perhaps more interesting to speculate is that this 
small difference records the anticipated tiny contribution that 
$\Delta R\neq0$ would make to the \hip photometry, but not our LDRs which are 
more directly sensitive to temperature changes. Our evidence for 
this is too slim for a credible claim but the idea might be successfully 
explored with more suitable data.

This completes our work with LDRs. It confirms past conclusions: \nuoa 
continues to have a very thermally-stable photosphere, which is difficult 
to reconcile with surface variability as we presently understand its 
many variations. As discussed in considerable detail in the LDR work reported 
in R15, neither star spots nor pulsations are believable scenarios for creating 
the planet-like RVs in these circumstances. The same explanations there apply 
here.

\bfi
\bc
\rotatebox{-90}{\scalebox{0.32}{\includegraphics{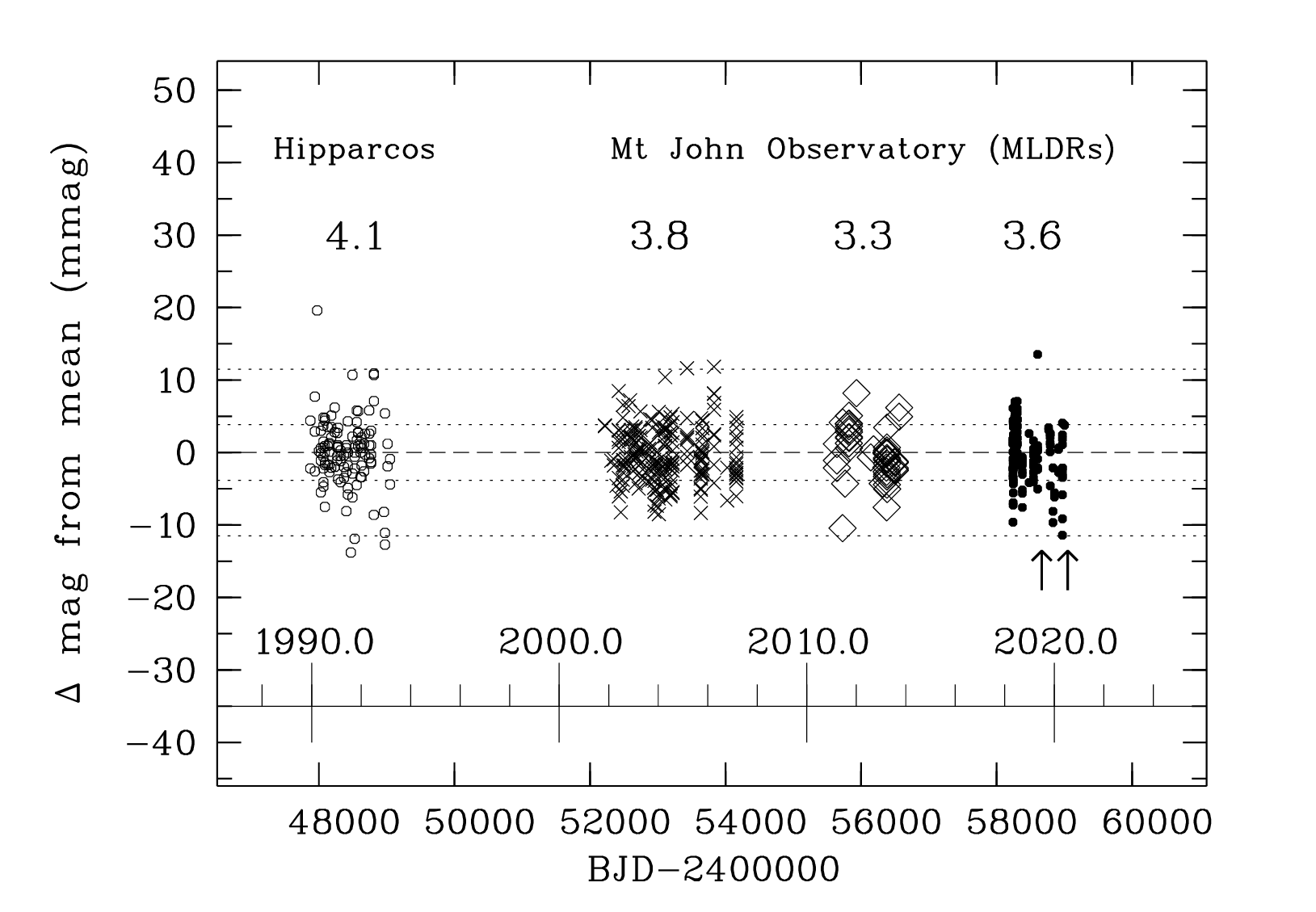}}}
\caption{The 398 magnitude differences, \mldr, derived using \eql{stefanb} and 
compared to \mhip from 116 best-quality ($\rm flag = 0$) 
\hip observations. The value above each series is the standard deviation of 
\dmag (mmag) weighted by the standard errors. The upward arrows identify the 
midtimes of the two TESS photometric baselines (2019, 2020).}
\lab{hipptmldrs}
\ec
\efi

\su{{\it Gaia} and more clues negating a stellar origin for RVs}
\lab{sec:gaia}
With the benefit of {\it Gaia}'s parallax (see \tl{stellar}) and our 
confirmation of the photosphere's stable temperature, we can also review 
several parameters that have some relevance here. The {\it Gaia} parallax is 
about 10 per cent greater than the three values reported in R09, which includes 
that from \hip (ESA 1997). Consequently \nuoa's luminosity is less, which 
we now estimate is about 14\lsun (and $\mbv=+2.3$~mag; \tl{stellar}). Based on 
the weighted mean of the five temperatures given above (\ie $4825\pm21$~K), 
\nuoa's radius is also less \ie $5.3\pm0.4$\rsun 
(see R09, their eq.\,3). The star is therefore now classified more closely as a 
subgiant, \ie $\rm\sim K1\,IV$, and placed in a sparsely populated part of 
the H-R diagram (very near \gcepa), well-separated from the so-called 
instability strip and other known classes of pulsating stars. It remains to be 
seen if future revisions change this significantly.

Such a star is not known to have a rotation period anywhere near as long as 
that of the RV cycle \ie $\prv \sim 415$~d. In fact, the star's 
$v\sin i\sim 2$~\kms (see \tl{stellar}), our revised estimate for the radius, 
and assuming the rotational and orbital axes are parallel, \ie 
$\irot\sim71\dg$, predicts the star's rotation period is $\prot=125\pm30$~d. 
R09 devotes their sec.~4.1.7 to the considerable improbability that rotation 
could be significant for the RV signal's origin. This scenario was made even 
more difficult to believe when R16 doubled the time span of the persistent RV 
cycle to 12.5~yr, and is reduced even further by the new evidence presented 
here.

Finally, we assess if solar-like oscillations have any chance of influencing 
our observations of the now less-evolved \nuoa. One way is to apply the 
formulae from Kjeldsen \& 
Bedding (1995), specifically equations 7, 8 and 10 to our new values for 
$[\mathcal{R}, T_{\rm eff}, \mathcal{L}]$. Using the weighted mean for the 
stellar mass, $1.6\pm0.1\msun,$ such oscillations have a predicted velocity 
amplitude of about 2\ms, a maximum-power period of only 1.5~hr and a luminosity 
amplitude $\delta \mathcal{L}/\mathcal{L}\sim 54$~ppm at 6000\AA~ (the 
wavelength from Wein's Displacement Law for the peak luminosity for our mean 
\teff). These values are in excellent agreement with the graphical ones in 
Dupret \oo (2009) who studied both radial and non-radial oscillations. One of 
their stellar models is slightly evolved and similar to \nuoa (their Case A, 
fig.~5, panel~3 at $\nu=190\,\mu$Hz). Its non-radial modes have smaller 
amplitudes than the radial ones, and together these produce a low-amplitude and 
very complex frequency spectrum, nothing like \nuoa's planet-like RV signal. 
Such oscillations cannot have any significant impact on any of our data.

We now report our analysis of the chromosphere-activity indicators, the \caii H 
line and \hal, the study of which benefits from the long series of precise LDR 
results -- any photospheric contribution to these indices is almost certainly 
insignificantly variable.

\se{Chromosphere activity: \caii and \hal indices}
\lab{sect:chromo}
Many spectral lines have been identified as being useful for monitoring 
chromosphere activity in late-type stars (see \eg Lisogorsky, Jones \& Feng 
2019). One of the motivations for assessing such behaviour is the quest for 
increasingly precise RVs for exoplanet searches. Whilst many photosphere lines 
can be used as indicators for chromosphere 
activity, the classic ones with the strongest history are the resonance doublet 
lines of \caii H and K at 3968.47 and 3933.66\ang close to the boundary of 
the ultraviolet and visible (see \eg Wilson 1963; Linsky \& Avrett 1970; 
Wilson 1978; Duncan \oo 1991; Baliunas \oo 1995). However, these two lines have 
the weakness, which will significantly influence our study, that they are 
recorded in spectral regions that have inherently low \stn.

At the other end of the visible spectrum is \hal ($6562.808$\ang). This has 
the distinct advantages of much higher \stn than provided at \caii H and K 
($\sim4-5\times$ greater), and a better defined 
continuum without the complications of many neighbouring metal lines as \caii 
H and K also have. However, as we will show, \hal instead has the complication 
of telluric lines, and for a quiet star such as \nuoa, these lines 
dominate the line-index variability. The \stn differences resulted in our 
having a final set of about 200 \caii H-line indices 
but nearly six times as many \hal indices.
 
\su{Line index definitions including covariance}
\lab{sect:definitions}
An index is a conventional method to assess variability of a suitable line's 
core for evidence of stellar activity (for its earlier history see Griffin \& 
Redman 1960). Our index, $I$, includes the line's core interval \fzero 
and two reference intervals $F_1$ and $F_2$: 

\beq
I = \frac{2 \times \fzero}{F_1 + F_2}\ ,
\lab{index}
\eeq
\noi
where each flux interval $i$ comprises $N_i$ bins having relative fluxes $f_i$. 
Modern studies use either total fluxes or mean fluxes: the index derived from 
mean fluxes $\langle F_i\rangle$ exceeds that from total fluxes by $N_R/N_0$ 
(if $N_R=N_1=N_2$). If total fluxes are used, the error on each flux 
$\eps_i = \si_i \times \sqrt{N_i}$. If 
mean fluxes are used $\eps_i = \si_i/\sqrt{N_i}$, \ie the standard error on the 
mean, which ensures the index's relative error will be identical to that 
from total fluxes. We will always use mean fluxes.

\begin{figure*}
\bc
\rotatebox{-90}{\scalebox{0.54}{\includegraphics{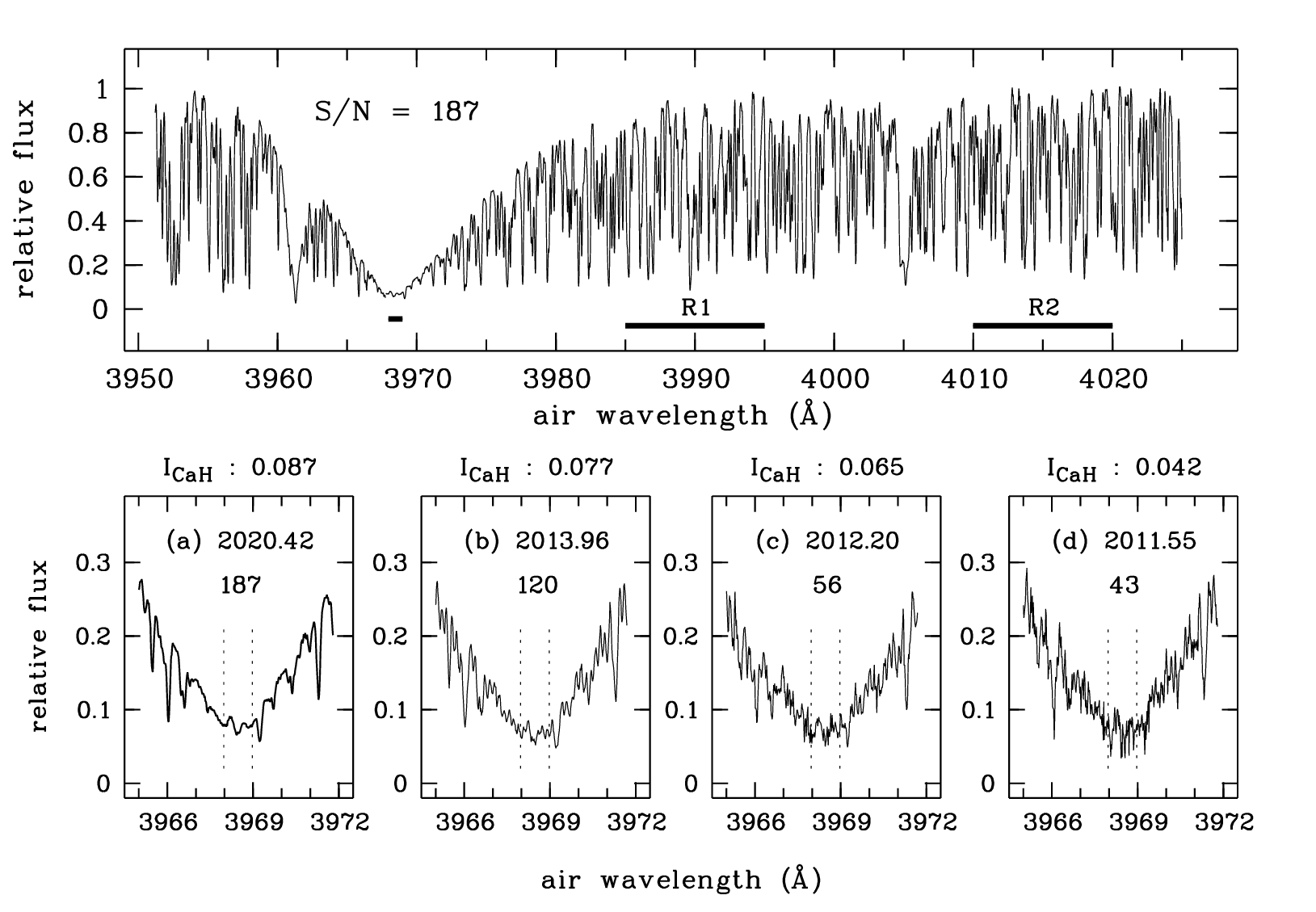}}}
\caption{Four \nuoa \caii H-line spectra for representative epochs, preliminary 
indices and \stn values. In each plot the core width represented is 1.0\ang. 
Top: The highest \stn spectrum across three orders. The core width and two 
reference intervals are marked with bold horizontal lines. The core is 
magnified below in (a). Below: The cores of four spectra with both decreasing 
\stn and index $I_{\rm CaH}$. Included are the $I_{\rm CaH}$, observation date, and 
\stn.}
\lab{CaHspectra}
\ec
\end{figure*}

The index error $\eps_{_{\rm I}}$ was calculated using error propagation:

\beq
\left(\frac{\eps_{_{\rm I}}}{I}\right)^2 = \left(\frac{\eps_{_0}}{F_0}\right)^2 + \frac{\eps_1^2 + \eps_2^2} {(F_1 + F_2)^2} = E_1 + E_2\ ,
\lab{properr}
\eeq
\noi
where we introduce the error symbols $E$ here and below for future 
reference. Whilst this is the typical calculation for such an index, 
\eql{properr} may be incomplete since the flux errors are calculated from 
the fluxes which, if suitably chosen, should be highly correlated. This 
suggests covariance deserves consideration.

The expression for the total covariance $CoV$ must be added to \eql{properr}:

\beq
CoV = - \frac{2\eps_{_{0}}\eps_{_{\rm T}}}{F_0(F_1+F_2)}\rho_{_{\rm 0,T}} + \frac{2\eps_1\eps_2}{(F_1+F_2)^2}\rho_{_{1,2}} = E_3 + E_4\ ,
\eeq
\noi
where $\rho$ is a correlation coefficient, $T$ represents the sum of the 
reference fluxes, and $\eps_{_{\rm T}}$ is calculated by also including its 
covariance term, i.e.,
 
\beq
\eps_{_{\rm T}} = \sqrt{\eps_1^2 + \eps_2^2 + 2\eps_1\eps_2\rho_{_{1,2}}}\ .
\eeq
\noi
The sign of $\rho_{_{\rm 0,T}}$ and $\rho_{_{1,2}}$ will determine if $E_3$ and/or 
$E_4$ increase or decrease the total error, or if $E_3+E_4\to 0$. If only one 
reference interval is used and covariance 
taken into consideration, the contribution of $E_4$ to the total error is 
absent.

\su{\caii H line spectra and data set}
\lab{sect:spectra}
When seventeen \caii H-line indices were reported in R16, the spectra we 
analyse here were available. However, most of the 700 archived \caii spectra, 
have quite low \stn ($\lesssim60$), and previously had been considered probably 
useless for any definitive study based on expectations and a preliminary 
sample's analysis.

As a definitive understanding of the \nuo system is yet to be attained, we took 
a second 
look at this large pool of spectra to review the earlier decision. We decided 
that it was a poor strategy to co-add spectra since the few available for most 
nights result in very little advantage and co-adding many across 
multiple epochs would provide fewer final indices, most obtained with greater 
complication and risk for errors as for instance Zechmeister \oo (2018) warns. 
Instead, a rather more novel approach was undertaken when a distinctive 
distribution that could be modelled revealed itself from a larger sample of 
indices.

The \hc archive provided an initial sample of 482 spectra. Many 
of the original 700 spectra could not be properly processed due to more 
prevalent cosmic rays and other complications. We restricted our analysis to 
the record of the \caii H line in $n=143$ as its \stn is 
about 40 per cent greater than its duplicate in $n=144$ and both records of 
\caii K. The 218 discarded spectra have $\stn\lesssim30$ at the H line 
record we used. Four H-line spectra of varying \stn are illustrated in 
\fl{CaHspectra} showing \nuoa's minimal chromospheric activity and the 
extent of slight infilling which varies primarily due to spectral noise.

\suu{Preliminary \caii indices and errors}
The origin of the dominant chromospheric flux is in the vicinity of the 
temperature minimum (Linsky \& Avrett 1970) which is presumably highly 
correlated with the LDR temperatures just described. The \caii H line 
has variable amounts of infilling and activity depending on, for instance, the 
star's luminosity (Wilson \& Bappu 1957). Typically the 
\caii H-line core of main-sequence stars is sampled with an interval of about 
1--1.1\ang (\eg Wilson 1978; Duncan \oo 1991; Cincunegui \oo 2007; Boisse \oo 
2009). The residual photospheric contribution which complicates accurate 
measurements of only the chromospheric component, for instance for surveys with 
different spectral types and luminosity classes (see \eg Oranje 1983; Rutten 
1984) can be neglected here since we are studying only \nuoa, which in any 
case has a very thermally-stable photosphere from evidence summarized in 
\fl{LDRperiodogram}.

Our H-line cores have a width of about 1.0\ang; see \fl{CaHspectra}. 
This detail provides further indirect evidence for \nuoa's revised less-evolved 
luminosity class (see \scl{sec:gaia}). Rather than commit to a 
single interval width $W$ we used a series in 0.1\ang steps from 0.9\ang to 
1.5\ang. We selected two reference intervals, R1 and R2, each 
10\ang wide, centered at $\lambda_{\rm air}=4015$\ang in order $n=142$ and at 
$\lambda_{\rm air}=3990$\ang in order $n=143$ (see top panel \fl{CaHspectra}). 
The second redder reference interval was chosen so that it would have a 
slightly higher \stn motivated by our mostly low-quality spectra.

Our preliminary 482 indices are shown in \fl{CaHindices}\,(a), 
given in relation to their \stn. These distributions fall into four ranges 
delineated by the vertical dotted lines in each panel at $\stn=30$, 40 and 60. 
In \fl{CaHindices}\,(a) the anticipated near constancy of the indices for a 
quiet star such as \nuoa is evident for $\stn\ge60$. The distributions of 
$E_1\ldots E_4$ and their sum are illustrated in \fl{CaHerrors}. Our errors 
yield $\rho_{_{\rm 0,T}}\sim-0.2$ and $\rho_{_{1,2}}\sim+0.7$, so that both $E_3$ and 
$E_4$ are positive, and $E_3$ is about as significant as $E_4$. $E_1$ is 
always the dominant term, but its contribution decreases as \stn 
increases. The error terms tend to plateau just above the threshold 
$\stn\sim60$, particularly for $E_3$ and $E_4$. Below this threshold, the total
error increases exponentially (\fl{CaHerrors}b), and 
above it, the minimum errors begin and average $0.0024\pm0.0002$. 
This error, incorporating the covariance terms, is equivalent to about 3.4 per 
cent of the index and about $2.5\times$ greater than the error without 
covariance included, a significant difference.

\bfi
\bc
\rotatebox{-90}{\scalebox{0.36}{\includegraphics{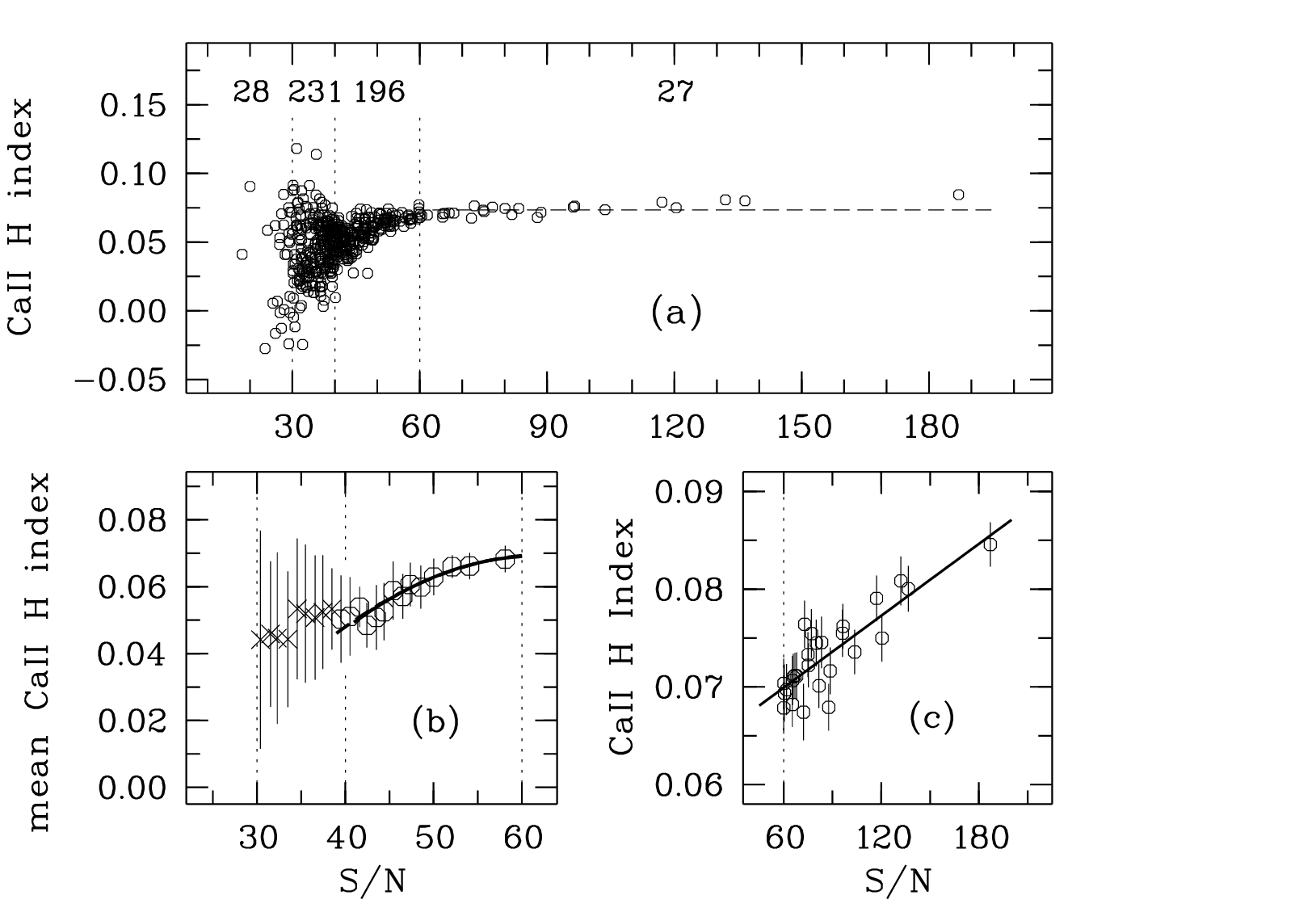}}}
\caption{The preliminary \caii H indices from 482 \nuoa spectra for a core 
width $W=1.0$\ang. (a) The indices, with the total for each subset bounded by 
$\stn=30,40,60$ given above. The dotted horizontal line identifies the 
weighted mean for the 27 spectra with $\stn\ge60$. (b) The weighted mean 
indices for each \mstn bin. The bold curve is the parabolic fit to the 13 \mstn 
bins in the range [39, 60]. (c) The linear fit for $\stn\ge60$. In (b) and (c) 
the y-scale is magnified for clarity and the vertical solid lines represent 
$\pm1\sigma$.}
\lab{CaHindices}
\ec
\efi

\bfi
\bc
{\scalebox{0.37}{\includegraphics{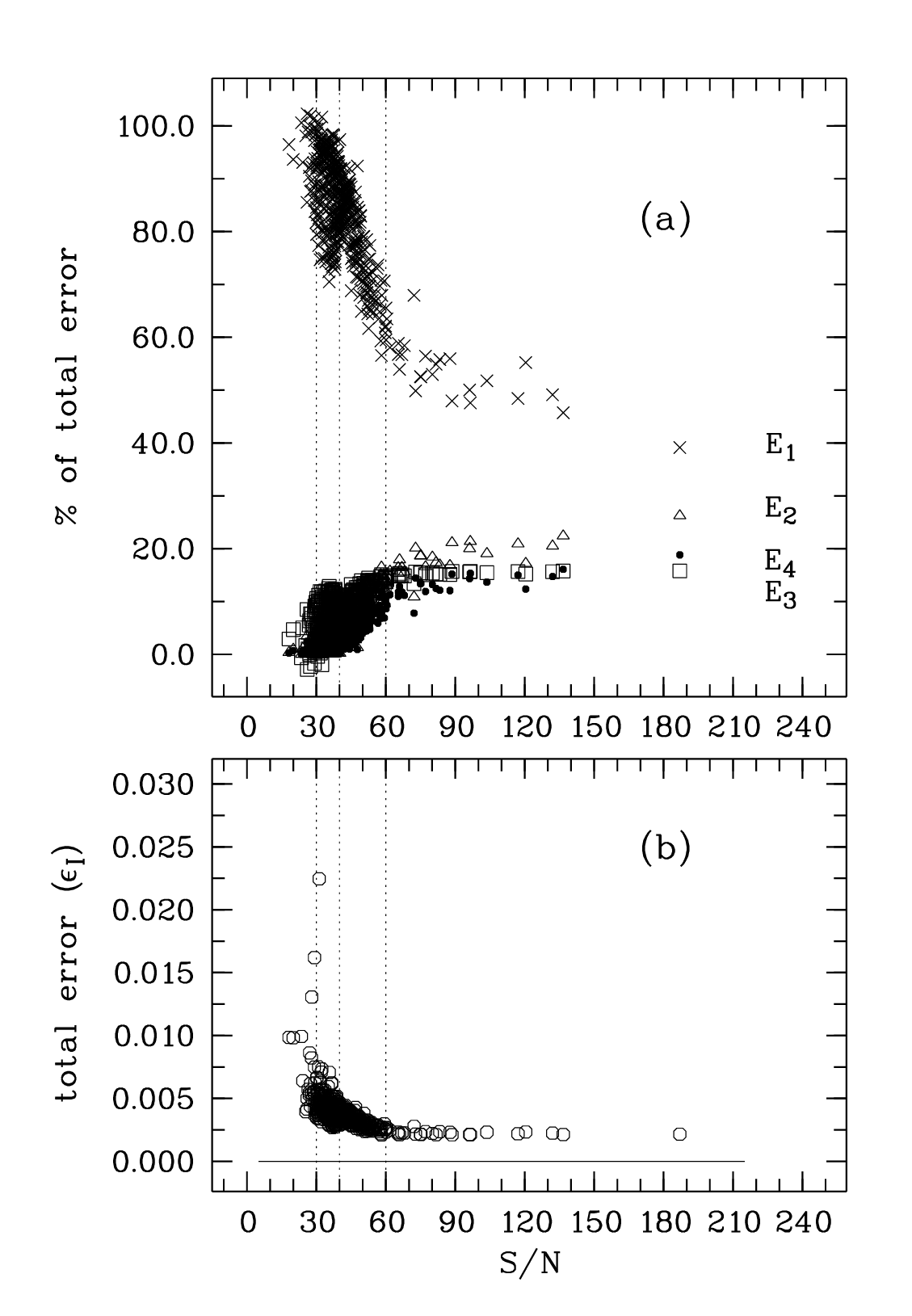}}}
\caption{The preliminary \caii H index errors from 482 \nuoa spectra. (a) The 
percentage fraction each error term contributes to the total. The single 
isolated symbols label the values from mid-2020 (JD245\,9002). The three dotted 
vertical lines are as in \fl{CaHindices}. The errors are for a core width 
$W=1.5$\ang since narrower widths have $E_2$, $E_3$ and $E_4$ increasingly 
coincident, approaching a mutual minimum contribution of about $10-15$\%. (b) 
The total errors given by the sum $E_1 + E_2 + E_3 + E_4$.}
\lab{CaHerrors}
\ec
\efi

\suu{Modelling the trends in \fl{CaHindices}\,(b) and (c)}
\lab{sect:tparabolic}
The apparently near-constant 27 indices with $\stn\ge60$ in 
\fl{CaHindices}\,(a) in fact delineate a fairly well-defined sloping line as 
shown below in \fl{CaHindices}\,(c), even without the high-\stn anchor. This we 
modelled with a linear fit weighting each index by $1/\eps_{_{\rm I}}^2$ to give 
our modified indices:

\beq
I^\prime = I - (a_1S + a_0) + \langle I_{60+}\rangle \ ,
\lab{linfit}
\eeq
\noi
where $S$ represents \stn and $\langle I_{60+}\rangle$ is the weighted mean of 
these 27 indices and defines our zero-point for $I^\prime$. Note that if 
our spectra had \stn limited to [60, 90] the implied slope of the distribution 
$\stn\ge60$ would disappear. The modelled indices have larger errors determined 
by the rescaled original index error created by the fitting process, the 
RMS of the linear fit (which varies with the core width of the H line), 
and the standard deviation of the original mean 
$\si_{\langle I_{60+}\rangle} = 0.004$, each added in quadrature. 

It is impossible to guess what the average behaviour of the 427 indices is in 
the interval [30, 60] in \fl{CaHindices}\,(a), so we created a series of narrow 
\stn bins, typically with $\Delta\stn = 1$. Starting at $\stn=30$ we extended 
the $\Delta\stn$ range in one unit steps until we had at least ten indices in 
each bin. This provided 23 bins. For each bin we calculated the average and 
standard deviation of the \stn and the indices, each index weighted by its 
error as above. These statistics, in relation to each bin's mean \stn are 
plotted in \fl{CaHindices}\,(b).

For $\stn\ge39$ the indices 
are adequately fit by a parabola. This fit though requires further care as it 
is not obvious what is its lower limit, which we label as \snth. Its choice is 
another balancing act between precision and sample size for our final time 
series since the binned indices have increasingly higher standard deviations as 
\stn decreases. We discarded 221 spectra in ten bins with $\stn<39$ as 
they have the largest standard deviations and are least consistent with our 
intended parabolic fit (marked with crosses `$\times$' in \fl{CaHindices}\,(b). 
For the 13 remaining $\langle\stn\rangle$ bins, which include 261 original 
indices, we calculated the model-fitted final indices and 
their errors for core widths of $W=[0.9, 1.2]$\ang. These four widths encompass 
those considered most suitable for our spectra (\ie 1.0 and 1.1\ang; see 
\fl{CaHspectra}).

\suu{Final indices and errors and their \redchi}
\lab{sect:chisqr}
Of the original 700 archived spectra, $\sim 60$ per cent are discarded 
from our complete analysis which somewhat vindicates the decision several years 
ago. Our errors and indices are now final. The errors each comprise two 
conventional terms $E_1$ and $E_2$, two covariance terms $E_3$ and $E_4$ and 
three terms from each model depending on the spectrum's \stn, making a total of 
seven terms. Each error is approximately doubled by the model-fitting 
calculations.

We calculated \redchi for our final indices and errors for our grid of four 
H-line core widths and thirteen \snth values. \fl{chisqrplots} shows that our 
atypical analysis of mostly low-\stn spectra and the inclusion of covariance 
ultimately provides strong evidence for the consistency of our indices, errors 
and models. For example, for the H-line core width $W=1.0$\ang, 
$\redchi\approx 1.0$ for the four consecutive parabolic fit thresholds 
$\snth=[41, 44]$. Without covariance $\redchi=2$. Our final relative errors 
average about seven per cent for the 198 indices defined by $\snth=41$. In 
\fl{adjustedCaH} we illustrate our final indices and errors for the time 
series defined by $\snth=41$ and $W=1.0$\ang which corresponds to 
$\redchi=0.99$. This example is approximately normally distributed as the lower 
two panels illustrate. We also derived $\redchi=0.99$ for the 27 indices from 
the spectra with $\stn>60$.

\bfi
\bc
\rotatebox{-90}{\scalebox{0.31}{\includegraphics{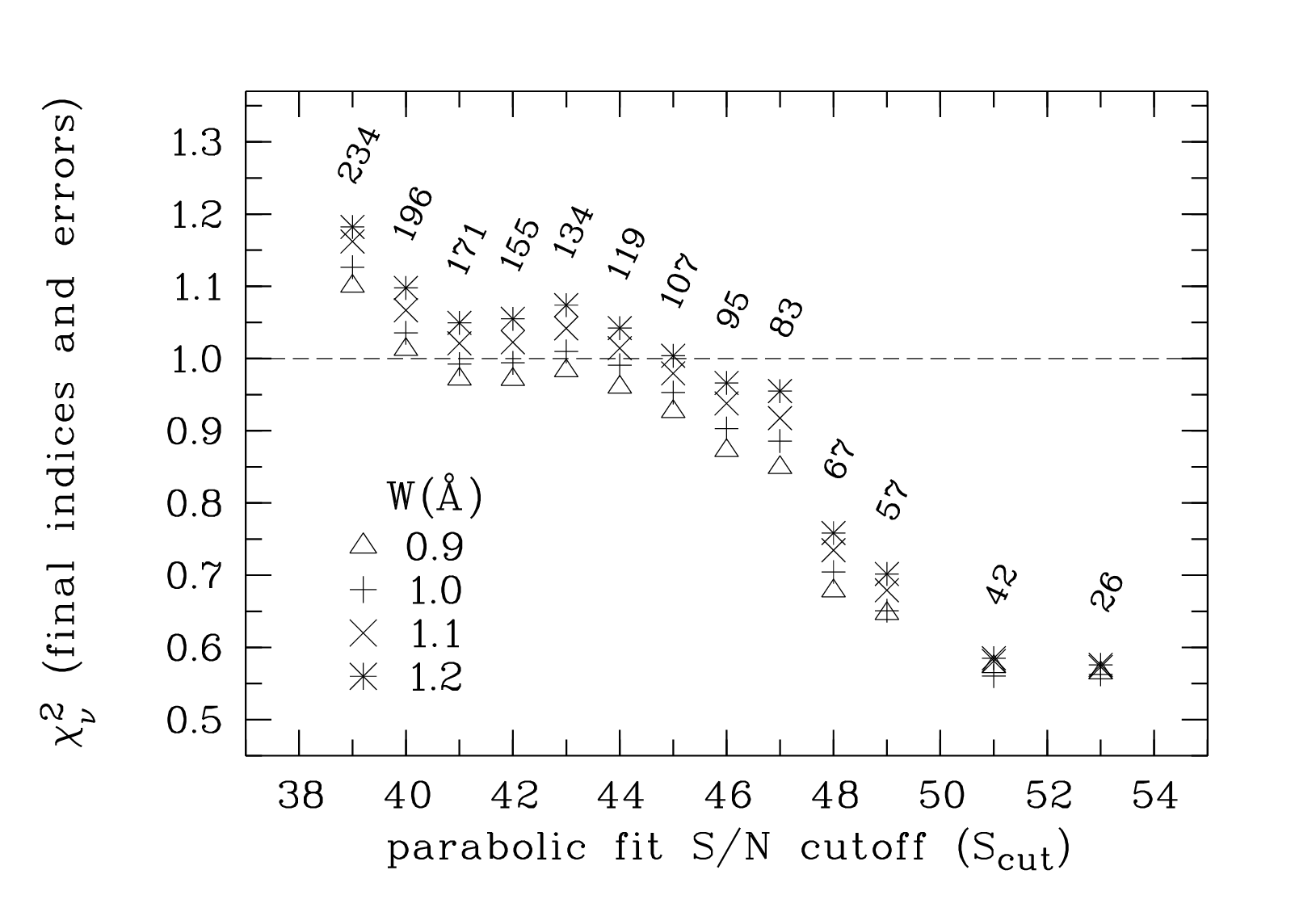}}}
\caption{\redchi values from our final indices and errors in terms of \snth. 
The results for four \caii H-line core widths $W$ are shown. The number above 
each \snth is the total number of indices included in each parabolic fit 
(shown in \fl{CaHindices}\,b), these finally combined with the 27 high-\stn 
indices.}
\lab{chisqrplots}
\ec
\efi

\bfi
\bc
{\scalebox{0.4}{\includegraphics{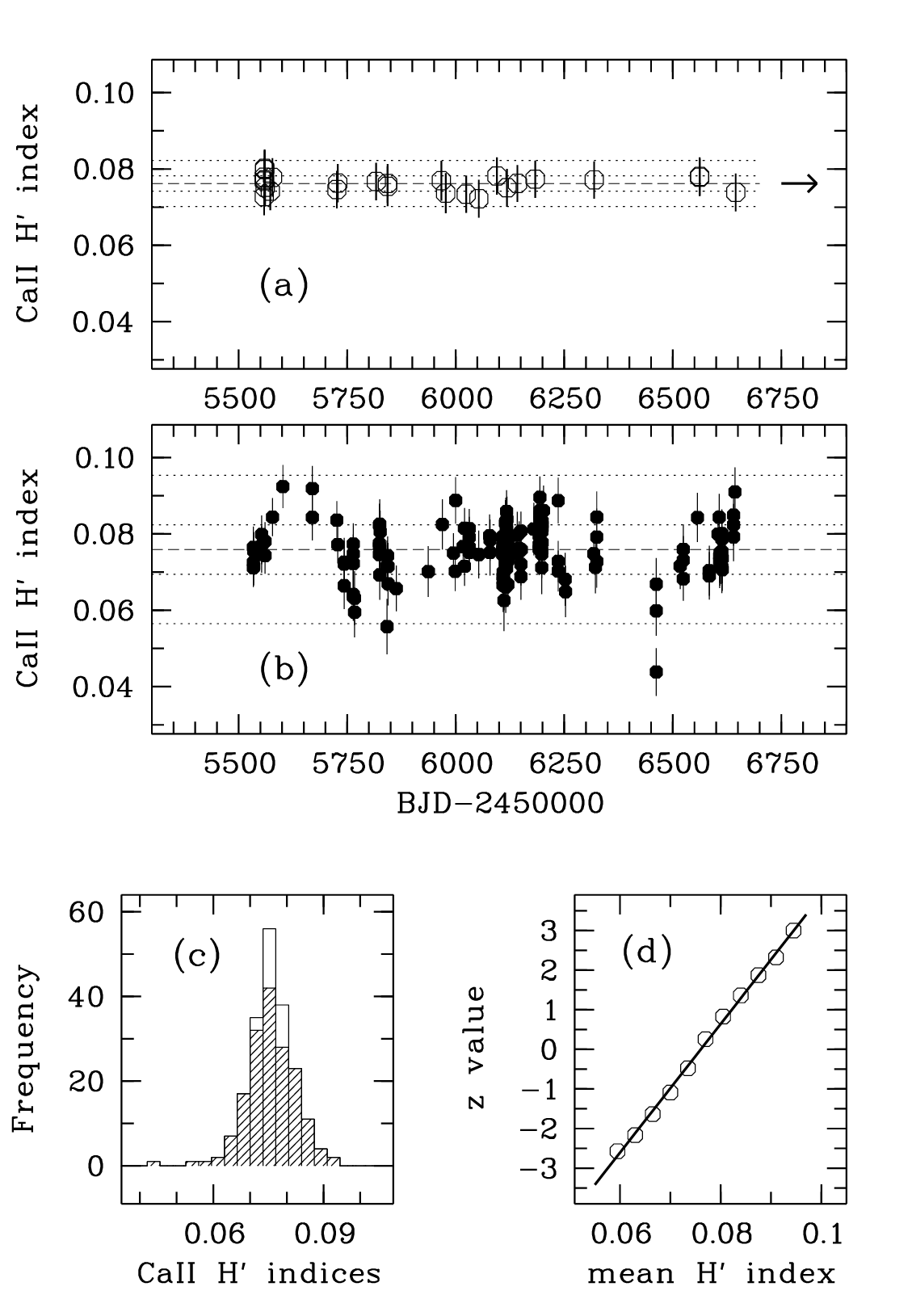}}}
\caption{Two distributions of indices based on the \caii H-line core width 
$W=1.0$\ang. Top panels: The dashed and dotted horizontal lines identify the 
mean and $\pm1$ and $\pm3\sigma$. The error bars are $\pm1\si$ and the y-scales 
are identical. (a) 27 indices with $\stn>60$, $\redchi=0.99$. The right arrow 
points to the mid-2020 (JD245\,9002) index. (b) 171 indices with $\stn\ge 41$, 
$\redchi=1.06$. (c) Frequency distribution of 198 indices. Open: (a) indices. 
Hashed: (b) indices. (d) Normal probability plot of 196 indices within 
$\pm3\si$ of the mean in (b), and the regression line based 
on the index mean and standard deviation.}
\lab{adjustedCaH}
\ec
\efi

\bfi
\bc
\rotatebox{-90}{\scalebox{0.27}{\includegraphics{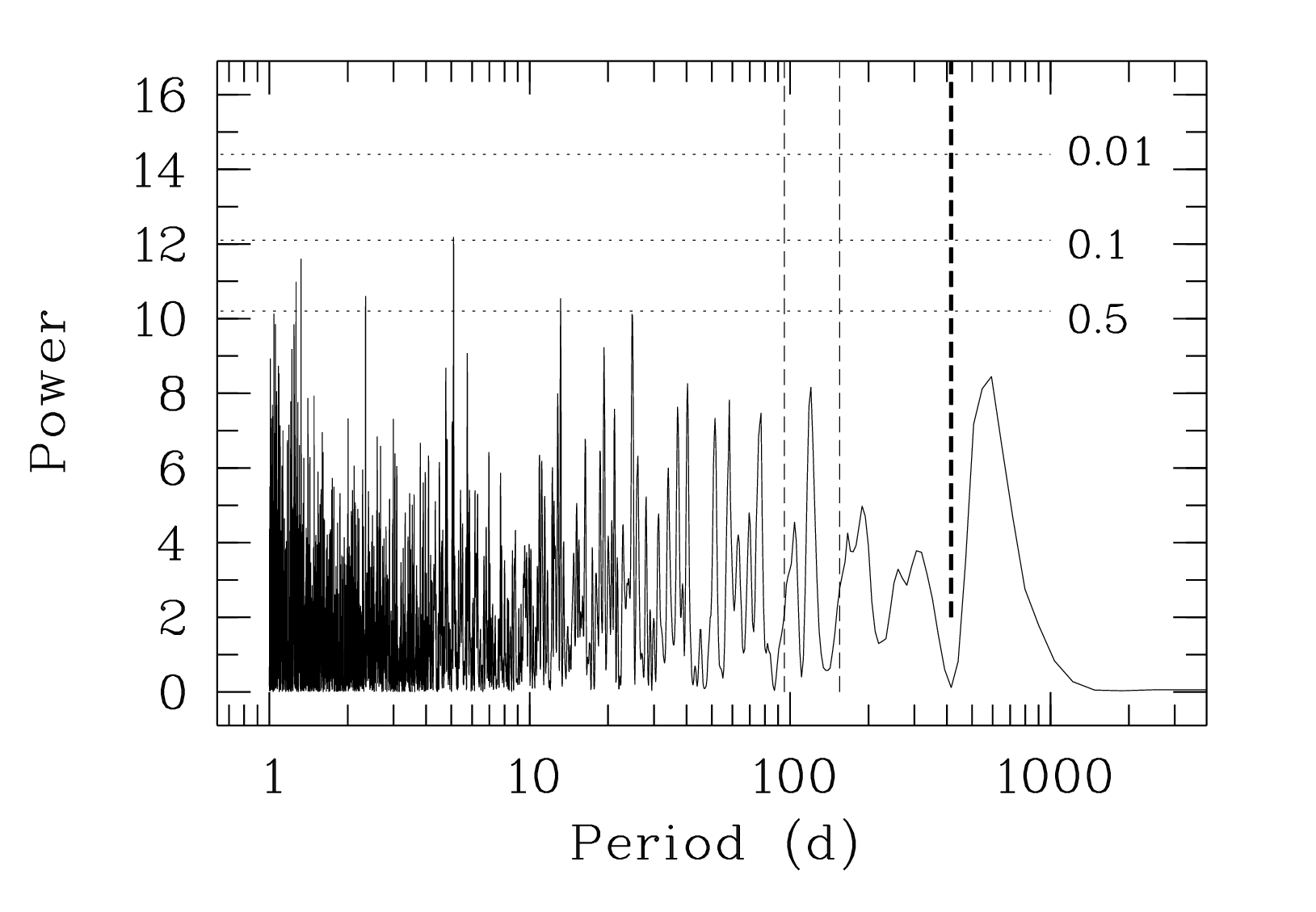}}}
\caption{GLS periodogram of the final series of 198 H-line indices using 
$\snth=41$, with the corresponding FAP levels. The pair of vertical lines 
identify the boundaries of our revised predicted $\pm1\si$ range of \nuoa's 
rotation period (see \scl{sec:gaia}) and the bolder line the planet-like RV 
period ($\prv \sim415$~d).}
\lab{CaHperiodogram}
\ec
\efi

As we found with our LDR work, within our errors, there is nothing here that 
suggests anything but quite randomly distributed data, regardless of the 
temptation to perhaps imagine some significantly periodic behaviour in 
\fl{adjustedCaH}\,(b) where removal of only a very few data points would make 
such a suspicion far less likely. Our GLS periodogram search as described in 
\scl{sect:ldrs} confirms this (see \fl{CaHperiodogram}). It was created using 
the indices plotted in \fl{adjustedCaH}. There is no significant peak 
corresponding to \nuoa's predicted rotation period ($\prot\sim125$~d) or the 
planet-like RV period ($\prv \sim415$~d) where a deep trough is in fact 
evident. Unfortunately, our lack of any detection of a credible rotation period 
(both here and with our LDR data) means we are still limited to 
estimates of \prot as in \tl{stellar} and \scl{sec:gaia} (where, in any case, 
we noted rotation remains a highly unlikely explanation for the RV cycle).

\su{\hal spectra and data set}
\nuoa's unique planet claim demands certainty about any conclusions 
relating to any index variability similar to the RV signal. Our \hal indices 
have this characteristic, having a quasi-periodicity in the 
vicinity of the conjectured planet's period. We intend to provide robust 
proof that this is caused solely by telluric lines.

\hal is a prominent photospheric absorption line in late-type stellar spectra. 
Its core can also include chromospheric activity. It was this profile 
variability that led to \hal being studied for $\beta$~Cephei in 1979 using one 
of the first digital detectors, an experimental Fairchild CCD (Young, Furenlid 
\& Snowden 1981).

We employed a long time series of \hal indices as densely sampled as possible, 
corresponding to the \iod-cell RVs reported in R16. Such a sample is critical 
for demonstrating the cause of the dominant variability in our indices. The 
1160 spectra archived over four years comprise 229 epochs. Thirty spectra (from 
18 epochs) were acquired without \iod which are useful to compare our indices 
with and without \iod involvement. For the majority with \iod lines, our \iod 
cell benefited from a temperature controller designed to minimise variability 
($50.0\pm0.1$~C), so that the density and extent of the \iod forest was assumed 
to be quite constant.\fn{We will discuss the absence of any significant impact 
of the \iod forest on our \hal indices in the final paragraph of 
\scl{sec:halpha}.} \hal is recorded in two \hc spectral orders, $n=86$ and 87. 
The \stn in order 87 is about twice that in 86 so we used only $n=87$, their 
range at \hal being $\stn\sim[90, 570]$ and their mean 220.

Our first evidence of the role of telluric lines is presented before indices 
are calculated. The high \stn region where \hal 
resides is also uncomplicated by large numbers 
of other photospheric lines such as the metal lines near \caii H and K. It is 
therefore very suitable for making a preliminary visual comparison of all of 
our spectra with respect to a reference. This has the advantage that the 
entire \hal region can be assessed for significant flux variations which may be 
concealed by the solitary line index. We ensured all our spectra have an 
essentially level continuum with a maximum relative flux $f_{\rm max}=1.0$, and 
that each \hal core centre is measured accurately to within a pixel by also 
comparing the fitted centre of the sharp \cai line ($\sim 6572.8$\ang) redward 
of \hal with each \hal core fit. We calculated the relative flux differences 
$\Delta f\di$ between each spectrum and our reference (see \fl{allcores}).

\bfi
\bc
\rotatebox{-90}{\scalebox{0.3}{\includegraphics{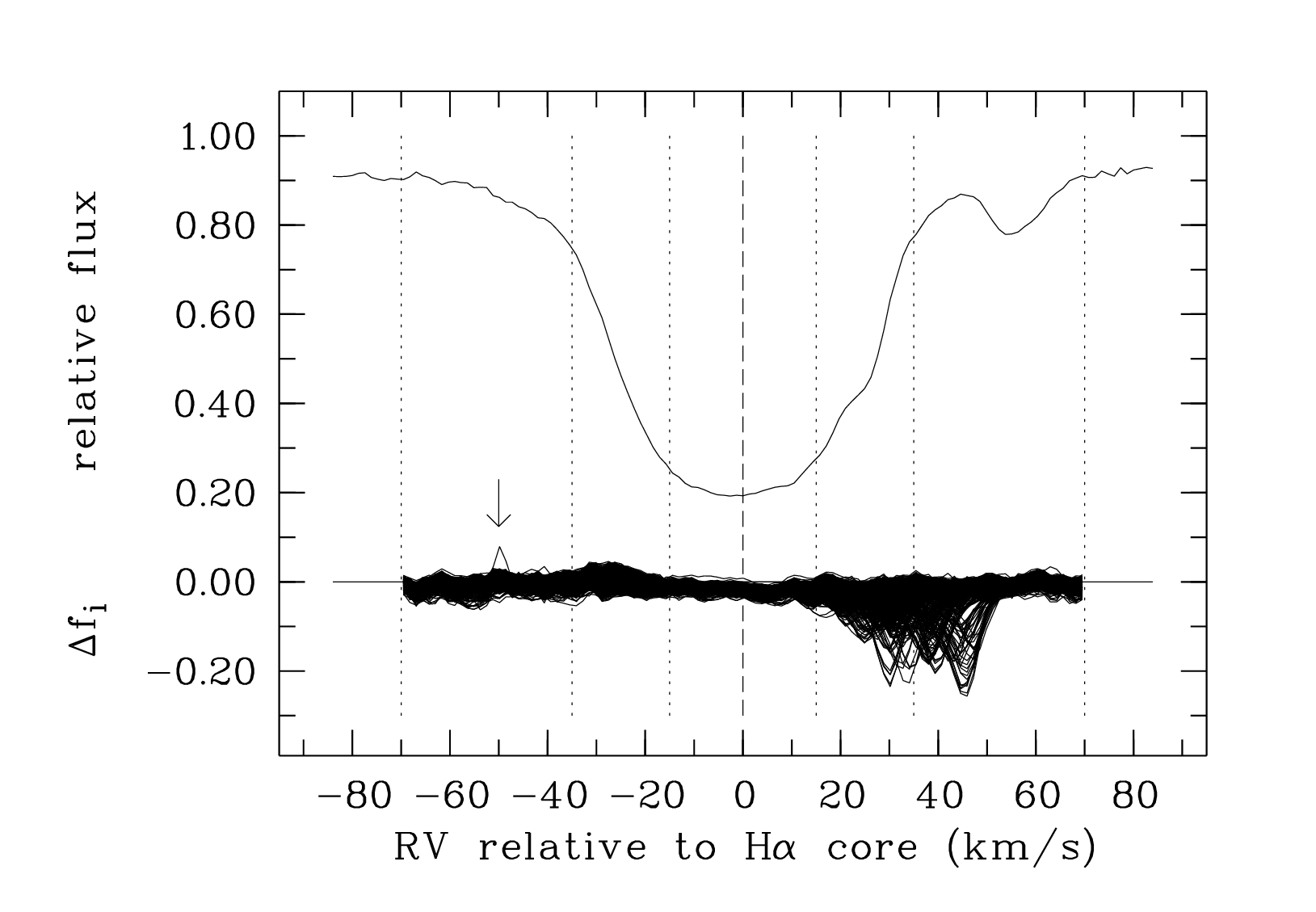}}}
\caption{1160 \hal flux-difference curves ($\Delta f\di$) relative to the 
\nuoa reference spectrum plotted above. The core boundaries of our \hal indices 
are identified with dotted vertical lines 
at $[-15,+15]$, $[-35,+35]$, and $[-70,+70]~\kms$. The small arrowed feature is 
likely to be due to a cosmic ray since it does not appear in the 
duplicate record in order 86.}
\lab{allcores}
\ec
\efi

This plot alone is strong evidence that \hal has very little core activity 
since the $\Delta f\di$ variations for the core interval $[-15,+15]~\kms$ are 
actually the least of the wide range shown. Instead we see time-dependent 
line movement contaminating both shoulders of \hal, but particularly the 
redward one, which subsequent evidence will prove are telluric lines. We 
were fortunate that the strongest telluric lines did not engulf our line cores 
as may occur in other circumstances.

\suu{Non-stellar lines and \nuoa's absolute RV}
All of our spectra include telluric lines and most also the \iod forest. An 
example of these non-stellar lines in a \hc spectrum of the fast-rotating Be 
star Achernar ($\alpha$~Eri, B6V, $v\sin i\sim 250$~\kms) is provided in 
\fl{tellI2}. About a dozen prominent telluric lines are recorded but the \hal 
region includes many more fainter ones.\fn{For instance, within the 95\ang 
shown in \fl{tellI2}, the online database HITRANonline (https://hitran.org) 
lists 576 water vapour lines, the principal species contributing to the 
telluric spectrum. For details of its latest release see Gordon \oo (2017).}

\bfi
\bc
\rotatebox{-90}{\scalebox{0.3}{\includegraphics{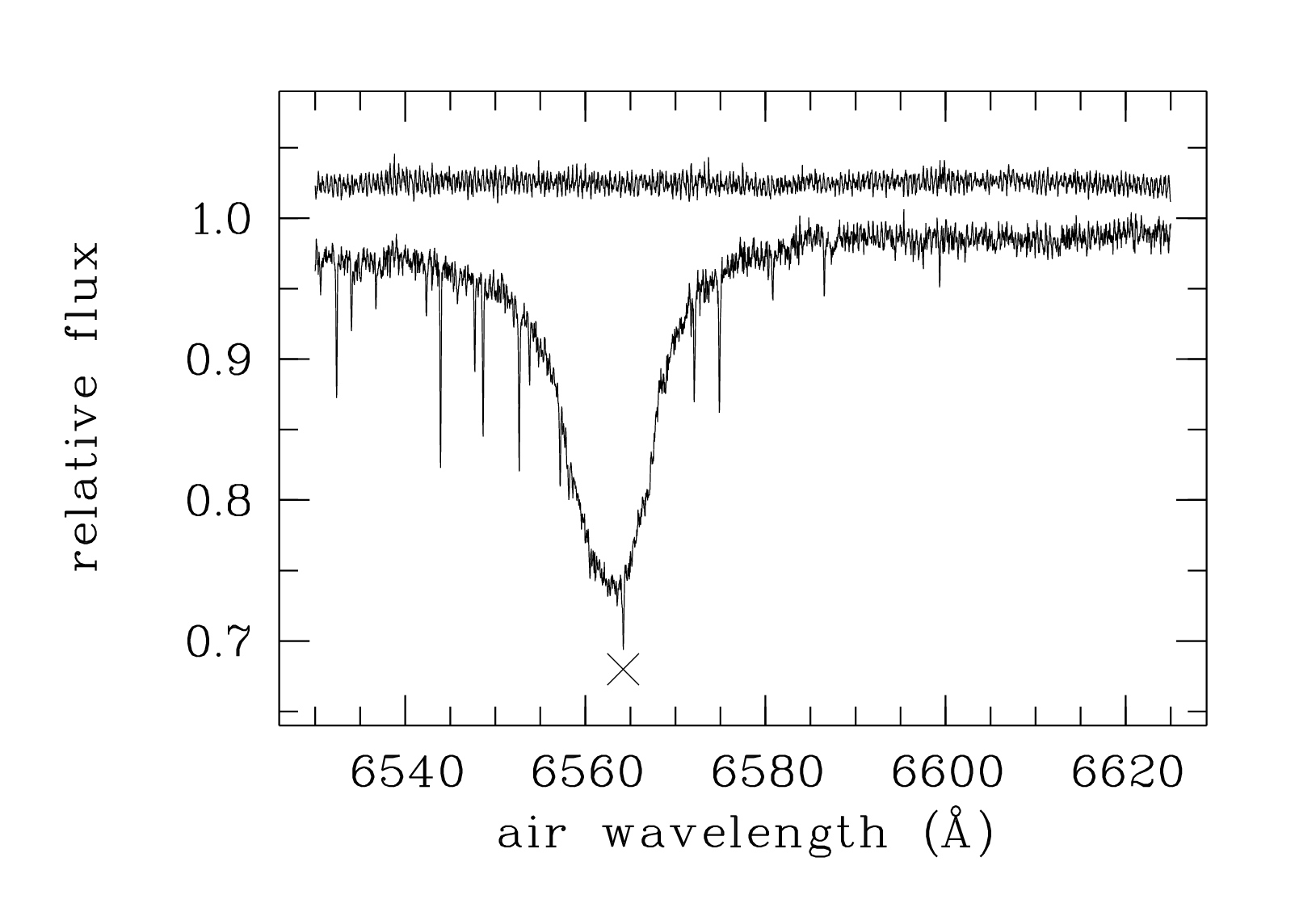}}}
\caption{Above: the weak but dense \iod-line 
forest in the vicinity of \hal (from a white+\iod spectrum). Below: A spectrum 
of the fast-rotating $\alpha$~Eri (B6V, $v\sin i\sim 250~\kms$) including 
\iod and about 15 more prominent telluric lines. The wavelength range is about 
$25\times$ greater than the equivalent of the RV scale in \fl{allcores}. The 
telluric line marked with `$\times$' ($\sim 6564.2$~\ang) also dominates 
the redward shoulder of the $\Delta f\di$ curves in \fl{allcores}.}
\lab{tellI2}
\ec
\efi

\nuoa's absolute RV is intimately connected to the relative behaviour of 
its lines and any non-stellar lines. We measured the absolute radial 
velocity (\rvcai) of \nuoa using the shift from the rest wavelength of the \cai 
line at $\sim6572.8$\ang. This line is recognized for its relative stability 
(\eg K{\"u}rster \oo 2003; Robertson \oo 2013), and for our purposes, the 
resulting low-precision RVs are adequate. Our \rvcai estimates and the 
best-fitting curve 
are provided in \fl{rvcurve}. Vertical lines identify the RV minima which have 
different time spans $\Delta t$ between them making them ideal for connecting 
the \rvcai variations to our index variations that we will next describe. Note 
that all of the $\Delta t$ intervals differ from the orbital period of the 
conjectured planet and the predicted rotation period of \nuoa, and critically, 
the $\Delta t$ differences are due to \nuoa's orbital RVs that a single star 
would not provide.

\bfi
\bc
\rotatebox{-90}{\scalebox{0.3}{\includegraphics{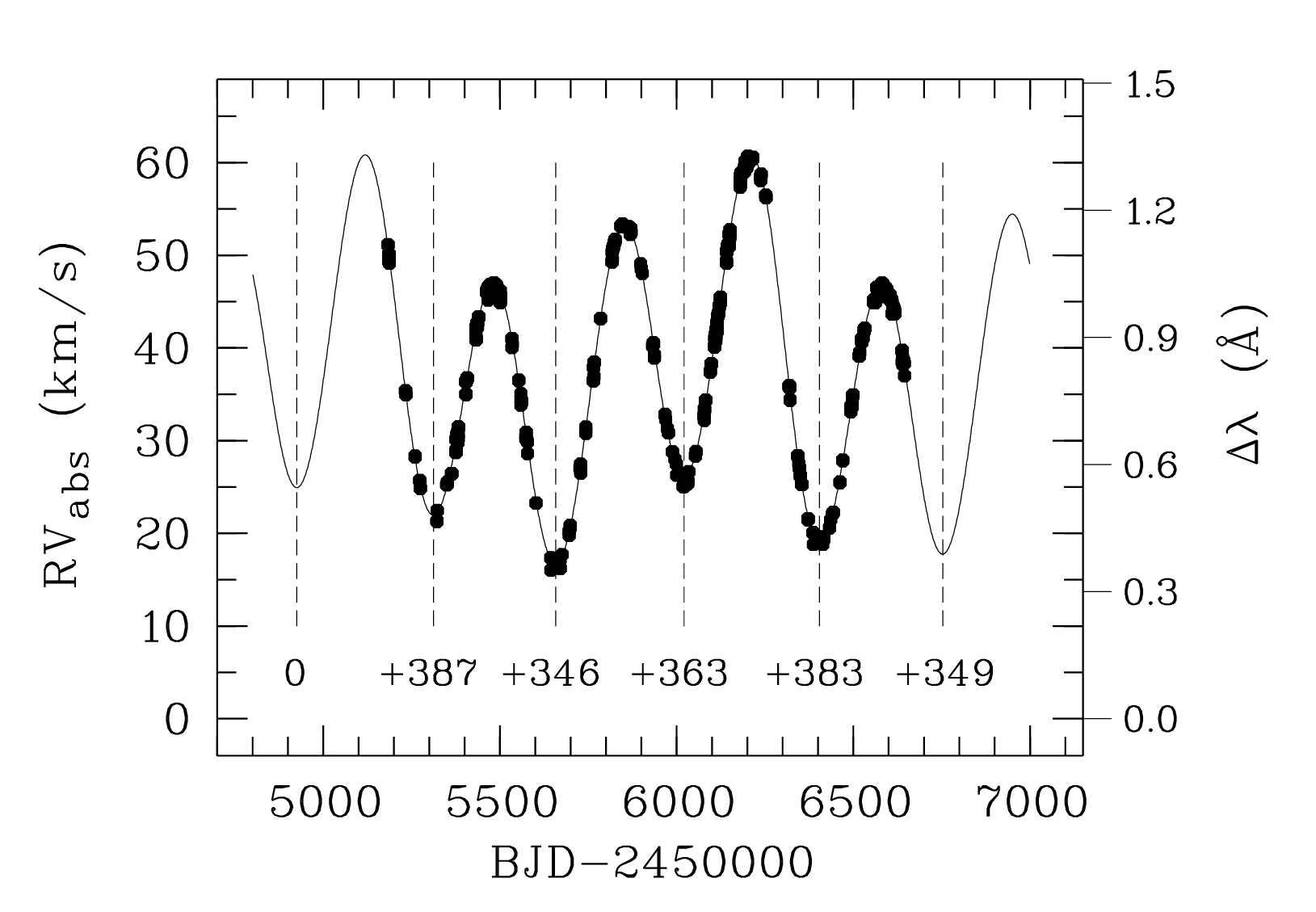}}}
\caption{The absolute radial velocity of 1160 \nuoa spectra based on the 
wavelength shift $\Delta \lambda$ of the \cai line at 
$\lambda_{\rm air} = 6572.8$\ang. The vertical dashed lines mark successive 
\rvcai minima with the time interval $\Delta t$ to the integer day between them 
(their average is 365.6~d). The curve is based on the barycentric RV 
corrections ($\sim[-16, +16]~\kms$) and the orbital solutions for \nuoa and the 
conjectured planet (Ramm \oo 2016), and our best-fitting estimate of the 
systemic RV ($+37.4$~\kms).}
\lab{rvcurve}
\ec
\efi

\fl{halphaCaI} includes the \hal spectrum with the highest \stn, a cartoon 
representation of the strongest telluric lines recorded in \fl{tellI2}, and 
identifies the \cai line. The reference intervals R1 and R2 are defined in the 
next section.

\begin{figure*}
\bc
\rotatebox{-90}{\scalebox{0.5}{\includegraphics{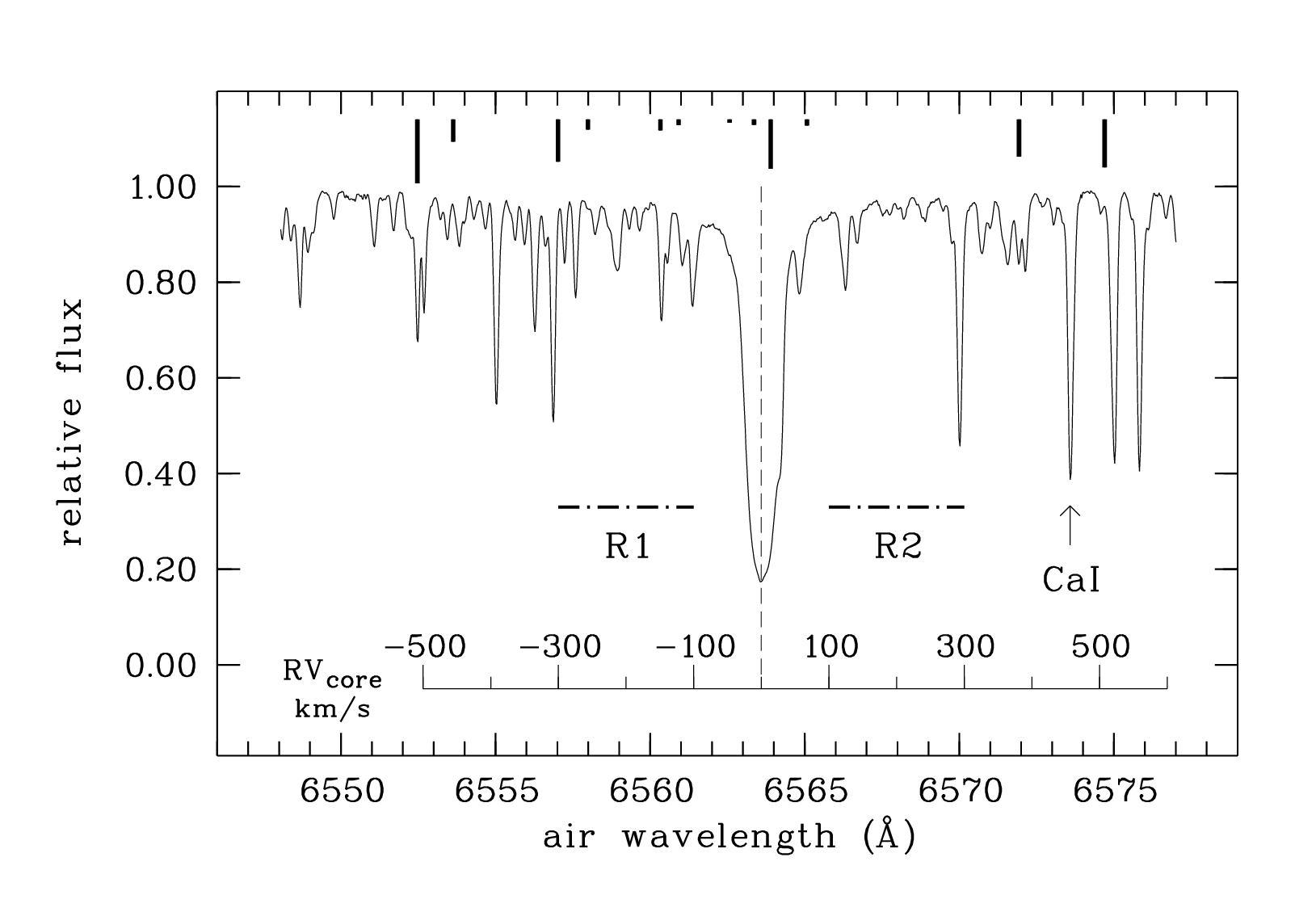}}}
\caption{The \hal region of our \nuoa spectrum with the highest $\stn\sim570$. 
It has $\rvcai\sim +35.9$~\kms which coincidentally is about midway in the 
range of RV values depicted in \fl{rvcurve}. The core centre is marked 
with a vertical dashed line at $RV_{\rm core}=0$ and our two reference intervals 
R1 and R2 located accordingly. The stellar spectrum and $RV_{\rm core}$ scale 
will move relative to the wavelength scale due to \rvcai, whereas the cartoon 
of telluric lines above the spectrum is fixed. Note this spectrum's \cai line 
is bounded by two strong telluric lines separated by $\sim130$~\kms and is 
discussed in \scl{sect:CaI}.} 
\lab{halphaCaI}
\ec
\end{figure*}

\suu{\hal indices}
\lab{sec:halpha}
The literature describes many choices for the flux intervals for \hal indices. 
For a survey of their variety see \eg K{\"u}rster \oo (2003); Cincunegui \oo 
(2007); Gomes da Silva \oo (2011); Ortiz \oo (2016); Zechmeister \oo (2018). We 
used three \hal core widths, with the centre defining the RV zero-point for all 
flux intervals. The narrowest, $[-15,+15]$~\kms, is suggested by the evidence 
in \fl{allcores}, and is similar to that used by K{\"u}rster \oo (2003). We 
also included the wider interval $[-35,+35]~\kms$ (similar to Zechmeister \oo 
2018) and one twice that, $[-70,+70]~\kms$, chosen to be just beyond the 
maximum absolute RV of \nuoa ($\sim61$~\kms). We used the two reference 
intervals R1 and R2 defined by Zechmeister \oo (2018), namely 
$[-300,-100]~\kms$ and $[+100,+300]~\kms$ and calculated the index errors 
including covariance as described earlier.\fn{We also 
calculated all of our indices using the reference intervals defined by 
K\"{u}rster \oo and Cincunegui \oo, which are distinctly different from 
Zechmeister \oo These index distributions were indistinguishable 
for the narrowest core width and barely so for the mid-width core. The indices 
we derived with the Zechmeister \oo reference intervals were significantly more 
precise for the widest core interval, which is why they are reported.}

Unlike our \caii indices, there is no correlation between \stn and our \hal 
indices ($\rho\sim+0.08$). However, none of the index distributions (see 
\fl{Zindices}) appear to be strictly 
random -- they all have evidence of cyclic behaviour with the dominant period 
of about one year, although this doesn't become better defined 
until the index uses the two wider \hal regions, the main reason for their 
inclusion.  Only the widest core width provides a significant maximum peak 
power in a periodogram and this is at 365.5~d (coincidently the mean of the 
approximated \rvcai timespans in \fl{rvcurve} is 365.6~d). This 
quasi-periodicity close to one year strongly implies the primary cause has a 
non-stellar origin, \ie telluric lines.

For the first time we demonstrate the usefulness of the five \rvcai minima 
lines as evidence that telluric lines are the dominant cause of our index 
variations: the set of lines has been offset in time by $-90$~d for the 
$[-35,+35]$~\kms core width but unadjusted for the widest one, and these 
included in \fl{Zindices} for the top two panels. Not only is adequate 
alignment found, simply made by eye, but the two offsets are unequal. The 
relative contributions of stellar and telluric lines vary if the flux intervals 
differ and this explains why the time shift to make this alignment is unequal 
for the two wider core widths that we can more easily assess -- in fact, for 
this reason, it is also more likely the shifts will differ. The more precise 
index variations of the widest core interval is perhaps explained by the more 
prominent roles of both the very stable photosphere and the additional telluric 
lines found here (which are also present in the reference intervals).

\bfi
\bc
{\scalebox{0.4}{\includegraphics{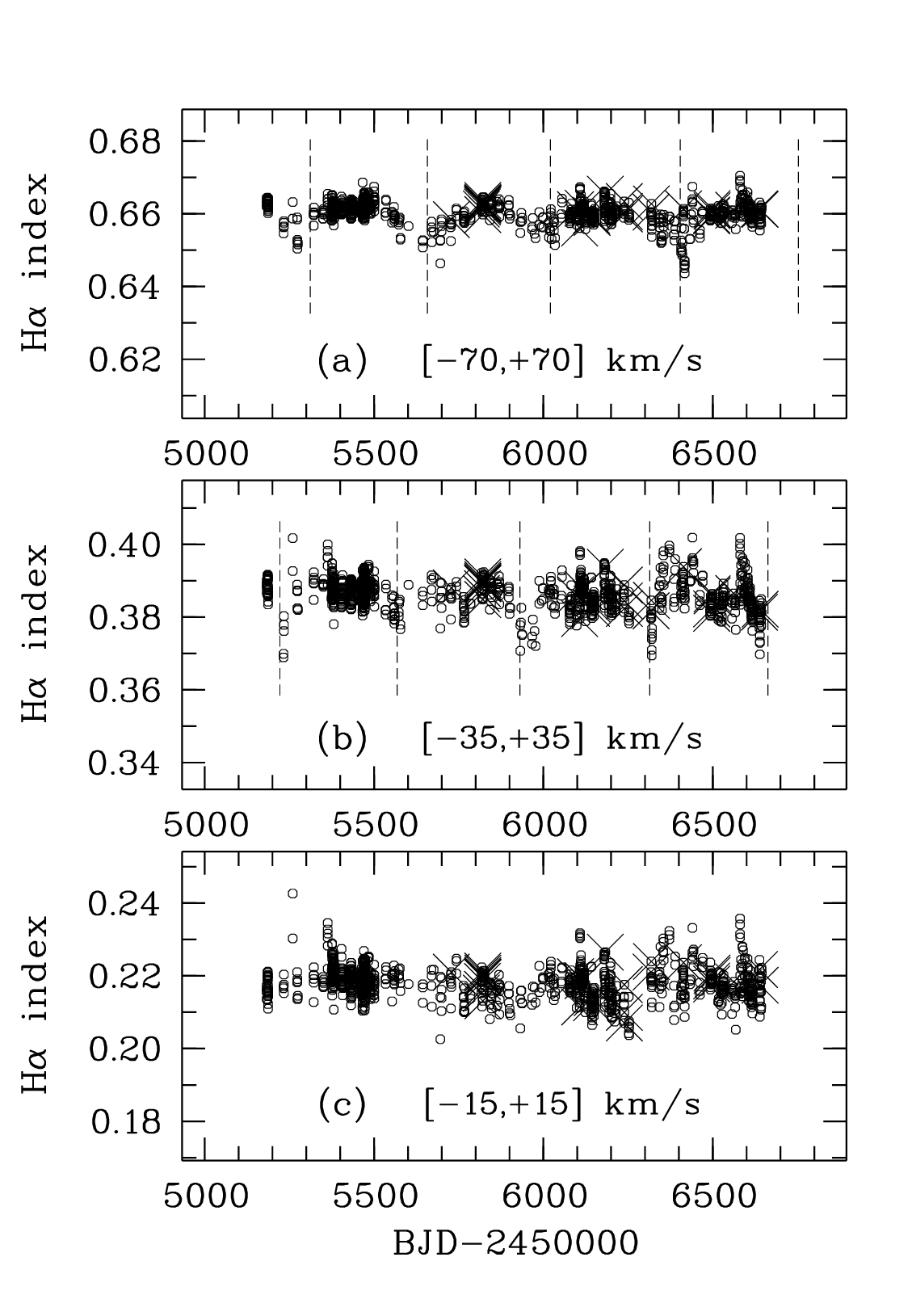}}}
\caption{\hal line indices from 1160 \nuoa spectra for three core intervals. 
`$\circ$': 1130 indices with \iod, `$\times$': 30 indices without \iod. (a) The 
dashed vertical lines are unshifted relative to their zero-point time in 
\fl{rvcurve}, but in (b) are shifted by $-90$~d, so these two 5-line sets are 
offset from each other. The y-axes all span the same range \ie 0.085. The 
mean percentage errors for the $\pm15$, $\pm35$ and $\pm70$~\kms indices are 
respectively $2.3\pm0.3, 6.9\pm0.2$ and $4.5\pm0.05$\,per cent.}
\lab{Zindices}
\ec
\efi

Thirty spectra were acquired without the \iod cell, the benefit of which 
is also clear from \fl{Zindices}: the no-\iod indices have a distribution that 
is consistent with those with \iod. To confirm this similarity, we also derived 
indices from our 135 new spectra (used for our LDR study) whose spectra also 
have no \iod lines. Their distribution is far less dense and uniform than our 
much larger series (which is the principal reason the former were not otherwise 
included for \hal) but are statistically indistinguishable from that derived 
from spectra with \iod. This demonstrates that the previous concern for using 
spectra including the \iod forest for an \hal study was unfounded, and 
duplicate the claim made in K\"{u}rster \oo (2003) that \iod involvement made 
no significant impact on their \hal and \cai indices for Barnard's star. That 
the complete absence or inclusion of \iod lines makes no significant difference 
to these indices also makes it clear that any temperature variations of the 
\iod cell leading to variability of the forest's extent is highly unlikely 
to make any difference either.

\suu{Other evidence}
\lab{sect:CaI}
Several more pieces of evidence ensure our \hal variations are dominated by 
telluric lines. Firstly, Bergmann (2015) used \hc to also measure \hal 
indices for $\tau$~Cet, $\delta$~Pav and $\alpha$~Cen~A and B in the same time 
span and frequently on the same nights as our \nuoa spectra. These indices use 
a wider core interval $[-115,+115]~\kms$ and different reference intervals but 
still those distributions and ours are very similar. Bergmann's four stars and 
\nuoa all have similar \rvcai ranges ($\sim 45-55$~\kms) so all had \hal moving 
relative to telluric lines in quite similar ways. Secondly, Bergmann's 
strategies to manage stray-light contamination between $\alpha$~Cen A \& B were 
strictly defined by his evidence his distributions were also substantially due 
to telluric line involvement, and more specifically, water vapour. Thirdly, 
Bergmann's reduction pipeline differed from ours so the chances of the two 
duplicating a reduction artefact is also highly unlikely. 

The final evidence comes from indices of the recognized `stable' \cai line used 
for our absolute RV estimates. We also calculated these using a series of core 
widths: the $[-15,+15]~\kms$ interval that is very similar to that used by 
K\"{u}rster \oo and Robertson \oo (2013) for their \cai indices, and a series 
in 10~\kms steps from 10--70~\kms. We found $[-30,+30]~\kms$ provided the least 
variation (half that provided by $[-15,+15]~\kms$). Again, the core interval 
$[-70,+70]~\kms$ recorded most distinctly the influence of telluric lines (see 
\fl{CaIindices}). Note that this wider interval is more than the separation of 
the two strong telluric lines bounding \cai in \fl{halphaCaI}, and so both 
influence some \cai indices. Our \rvcai minima lines are again successfully 
aligned to the dominant cycles for the wider core interval. These results show 
that 1.~the index distribution of the stable \cai line closely mimics our \hal 
indices and has the least variation of all our indices (the mean for 
$[-30,+30]$~\kms is $0.958\pm0.003 \equiv$~0.3 per cent), and 2.~if 
confirmation of stability is desired, \cai or any other line is best studied 
when less likely to be contaminated by tellurics.

\bfi
\bc
\rotatebox{-90}{\scalebox{0.34}{\includegraphics{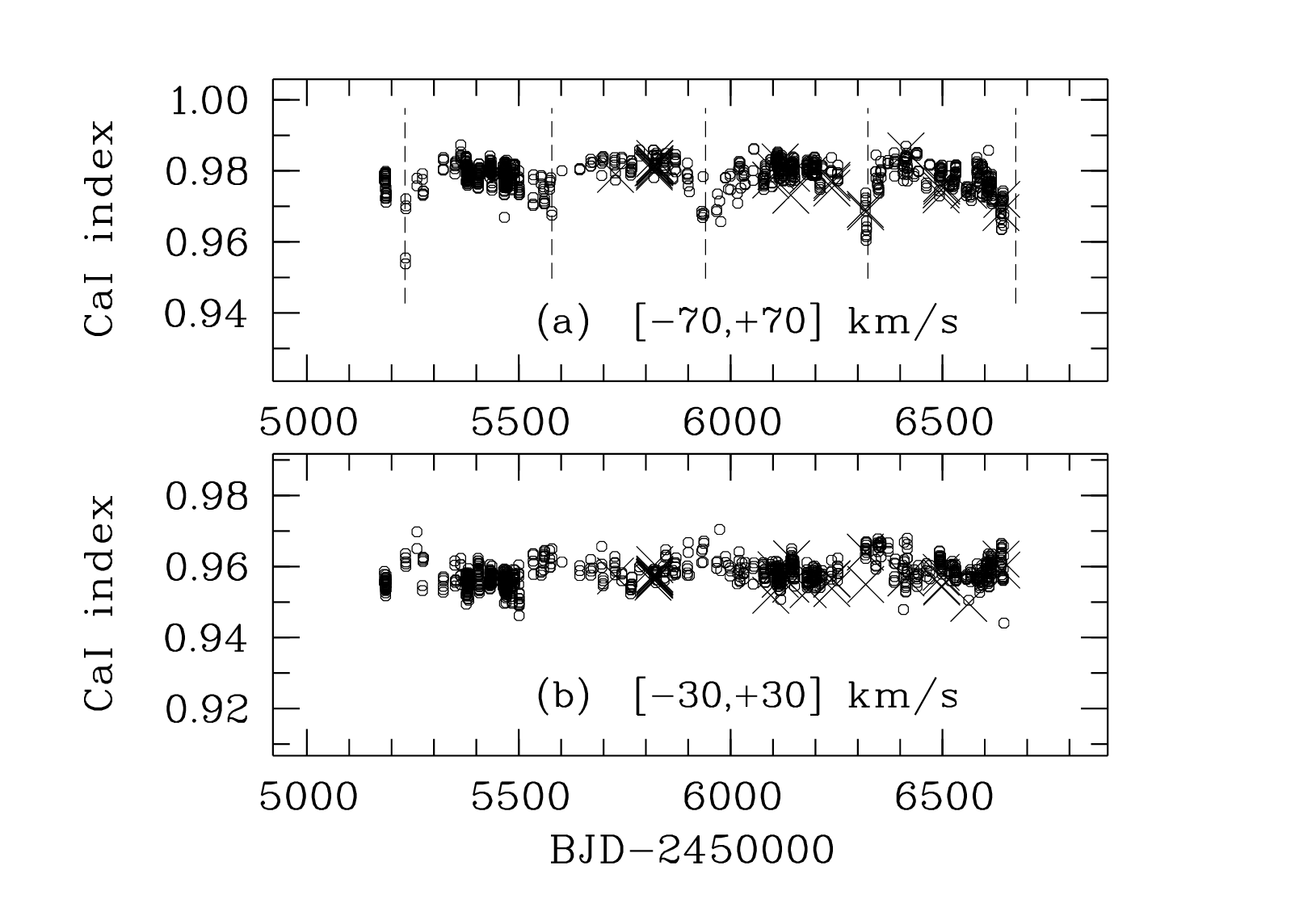}}}
\caption{\cai indices from 1160 \nuoa spectra for two core widths. 
`$\circ$': 1130 indices with \iod, `$\times$': 30 indices without \iod. The 
axis ranges are as in \fl{Zindices}. The dashed vertical lines in (a) are 
shifted relative to their zero-point in \fl{rvcurve} by $-80$~d.}
\lab{CaIindices}
\ec
\efi

\suu{\hal index errors and final comments}
We calculated our index errors $E_1\ldots E_4$ using the same methods 
described in \scl{sect:definitions}. Given the 
significant impact of the covariance terms $E_3$ and $E_4$ to our \caii index 
errors, it was somewhat surprising that they make little difference to 
our \hal errors, since the mean fluxes of the reference intervals are again 
highly correlated, just as they were for our \caii indices 
(\hal: $\rho_{_{1,2}}\sim0.9$). However, perhaps related to the influence 
of telluric lines, the associated errors ($\eps=\si/\sqrt{N}$) are not at all 
highly correlated, regardless of the core or reference intervals used, since we 
also checked this result with the K\"{u}rster \oo reference intervals. If this 
result is common in other \hal index studies it shows that the absence of the 
covariance terms also made no significant difference to their errors.

The core width that is most suitable for our original purpose -- assessing 
\hal's activity -- is the narrowest one, $[-15,+15]~\kms$. That index 
distribution is in the bottom panel of \fl{Zindices}. $E_1$ is again by far the 
dominant error term. The relative contributions of the four error terms to 
their total are (0.962, 0.046, 0.005, $-0.013$). The relative total errors have 
a mean $2.3\pm0.3$ per cent which yield $\redchi=0.9$. Both $E_3$ and $E_4$ are 
practically zero for two different reasons: $E_3\sim 0$ as 
$\rho_{_{\rm 0,T}}=-0.016$, and $E_4\sim 0$ primarily because the error portion is 
$\sim 0$ (and $\rho_{_{\rm 1,2}}=-0.30$, so that $E_4$ actually reduces the total 
error, but only insignificantly).

That we can precisely correlate the tiny effects of telluric lines in terms of 
\rvcai suggests our methods are reliable and any \hal activity is likely to be 
significantly less. Also, as a final minor detail, having established the 
origin of the dominant variations, we can now speculate that contributions to 
the almost nightly variations may be differences with the observations' 
airmasses and the humidity that the telluric lines' primary component, \ie 
water vapour, also records.

\se{Conclusion}
We report five new independent series of precise data that all indicate \nuoa 
is a very quiet star: two photometric series from TESS (2019, 2020), which more 
or less coincide with our LDR study of its photosphere (2018--2020), and our 
two large surveys of its chromosphere whose origin is the \iod-cell RV time 
series (2009--2013). The TESS photometry bounds the 1437 RVs (2001--2013) with 
\hip observations three decades before (1989--1993). A sixth line of new 
evidence is the {\it Gaia} parallax that indicates \nuoa is less luminous, and 
so 
more probably approximately K1\,IV. This also strongly favours the quiet-star 
classification, as does our review of solar-like oscillation scenarios. These 
results are consistent with all previous surveys, for instance, more recently, 
two bisector studies (R09 and R16), and two other LDR studies (R15 and R16).

There is no significant variability of any of these many photometric or 
spectroscopic time series, and certainly none correlated with the planet-like 
RVs. Other facts are against \nuo being a heirarchical triple-star system or 
the cause being related to \nuoa's rotation. Thus this overwhelming evidence 
demonstrates all non-planetary solutions continue to be implausible, which 
instead we interpret is only compatible with a retrograde planet whose 
properties are approximated in \tl{orbital}. 

With thousands of planets now described, \nuoab remains unprecedented in terms 
of the system's geometry. It still provides many opportunities for exploring 
unexpected dynamical models of planet formation and orbit stability, as well 
as exceptional observational opportunities for such a bright, short-period 
system's orbital behaviour. Recently found evidence that the secondary star may 
be a white dwarf was reported in the Introduction which would have a 
significant bearing on such studies. As more evidence accumulates only in 
favour of the planet, the flip-side of irrefutable evidence that the planet is 
instead an illusion is how unexpected any non-planetary alternative would 
then be.

We began by confirming how thermally-stable \nuoa's photosphere continues to 
be. This allowed us to confidently claim the photosphere, which has produced 
the enigmatic RVs for over a decade, is highly unlikely to contribute any 
variability to our chromosphere indices, nor be the source of the planet-like 
signal. Knowing the thermal stability of the photosphere is a useful advantage 
for any chromosphere study, a detail not commonly reported.

It is no doubt partly due to \nuoa's quiescence that has allowed our study of 
its \caii spectra with such low \stn and successfully pursue our atypical 
methods. It is also this stability that has helped us critically examine the 
telluric line contribution to our \hal spectra and quite clearly reveal its 
quasi-periodic behaviour and origin. All our results strongly imply they record 
mostly random processes and not something systematically cyclic.

Because similar studies use data that may be at least moderately correlated it 
seems that covariance deserves closer scrutiny in such cases. Even though 
the covariance terms have opposite signs, the correlation coefficients may 
still cause the covariance terms to increase the errors significantly -- as our 
\caii results demonstrate. Our methods for 
analysing low-\stn spectra may allow other so-far neglected archival 
material to be recovered usefully as we quite surprisingly achieved here. 
Finally, we introduced a simply described classification system for 
retrograde orbits that appears effective for our purposes, and may be 
of sufficient merit for use by others.

\section*{Acknowledgements}
We appreciate the anonymous referee's several suggestions that led to helpful 
improvements. DJR thanks M.\,F.\,Reid, then Physics \& Astronomy HoD, for 
renewing his Research Fellow status 
(2018--2020) allowing continuing access to academic and IT resources including 
the assistance of O.\,K.\,L.\,Petterson. DJR is also grateful to 
the UC Mt. John Observatory director K.\,R.\,Pollard and the present HoS 
R.\,Marquez for the generous allocation of observing time and research support 
for this project that provided us the opportunity to obtain more data for \nuo. 
We thank the observing technician F.\,Gunn for obtaining the spectra 
2018--2020 and K.\,R.\,Pollard and R.\,Marquez of the School of Physical and 
Chemical Sciences, University of Canterbury for sourcing observing and 
technical support for this project at the UC Mt John Observatory, including 
funding from The Brian Mason Trust, UC Foundation, The School of Physical and 
Chemical Sciences and The Otago Museum. DJR thanks M.\,K{\"u}rster (MPIA) and 
K.\,J.\,Moore (Christchurch) for helpful contributions. This research made use 
of \textsc{Lightkurve}, a Python package for Kepler and TESS data analysis 
(Lightkurve Collaboration, 2018).

\section*{Data Availability}
The data underlying this article will be shared on reasonable request to the 
corresponding author.

\bsp	% typesetting comment
\label{lastpage}

\begin{thebibliography}{}
\bibitem[\protect\citeauthoryear{Arkharov \oo}{2005}]{UVvar}
Arkharov~A.~A., Hagen-Thorn~E.~I., Ruban~E.~V., 2005, Astron.~Rep., 49, 526

\bibitem[\protect\citeauthoryear{Barnes \oo}{2020}]{barnes}
Barnes~J.~R. \oo, 2020, Nature Astronomy, 4, 41

\bibitem[\protect\citeauthoryear{Baliunas \oo}{1995}]{baliunas}
Baliunas~S.~L. \oo, 1995, ApJ, 438, 269

\bibitem[\protect\citeauthoryear{Bergmann}{2015}]{cbthesis}
Bergmann~C.~M., 2015, PhD thesis, Univ.~Canterbury

\bibitem[\protect\citeauthoryear{Bonavita \& Desidera}{2020}]{bona}
Bonavita~M., Desidera~S., 2020, Galaxies, 8, 16

\bibitem[\protect\citeauthoryear{Boisse \oo}{2009}]{boisse}
Boisse~I. \oo, 2009, A\&A, 495, 959

\bibitem[\protect\citeauthoryear{Campbell \oo}{1988}]{campbell}
Campbell~B., Walker~G.~A.~H., Yang~S., 1988, ApJ, 331, 902

\bibitem[\protect\citeauthoryear{Campanella}{2011}]{campanella}
Campanella~G., 2011, MNRAS, 418, 1028

\bibitem[\protect\citeauthoryear{Catalano \oo}{2002}]{catalano}
Catalano~S., Biazzo~K., Frasca~A., Marilli~E., 2002, A\&A, 394, 1009

\bibitem[\protect\citeauthoryear{Cincunegui \oo}{2007}]{cincunegui}
Cincunegui~C., D\'{i}az~R.~F., Mauas~P.~J.~D., 2007, A\&A, 469, 309

\bibitem[\protect\citeauthoryear{Duncan \oo}{1991}]{duncan}
Duncan~D.~K. \oo, 1991, ApJSS, 76, 383

\bibitem[\protect\citeauthoryear{Dupret \oo}{2009}]{dupret}
Dupret~M.-A. \oo, 2009, A\&A, 506, 57

\bibitem[\protect\citeauthoryear{Dvorak}{1986}]{dvorak}
Dvorak~R., 1986, A\&A, 167, 379

\bibitem[\protect\citeauthoryear{Eberle \& Cuntz}{2010}]{cuntz}
Eberle~J., Cuntz~M., 2010, ApJ, 721, L168

\bibitem[\protect\citeauthoryear{ESA}{1997}]{hip}
ESA, 1997, The Hipparcos and Tycho Catalogues, ESA SP-1200

\bibitem[\protect\citeauthoryear{Fuhrmann}{2004}]{fuhra}
Fuhrmann~K., 2004, AN, 325, 3

\bibitem[\protect\citeauthoryear{Fuhrmann}{2012}]{fuhrb}
Fuhrmann~K., Chini~R., 2012, ApJS, 203, 30

\bibitem[\protect\citeauthoryear{Brown \oo}{2018}]{gaia}
Gaia Collaboration, 2018, A\&A, 616, A1

\bibitem[\protect\citeauthoryear{Gomes da Silva \oo}{2011}]{gomes11}
Gomes da Silva~J., Santos~N.~C., Bonfils~X., Delfosse~X., Forveille~T., Udry~S., 2011, A\&A, 534, A30

\bibitem[\protect\citeauthoryear{Gordon \oo}{2017}]{hitran}
Gordon~I.~E. \oo, 2017, J. Quant. Spectrosc. Radiat. Transf., 203, 3

\bibitem[\protect\citeauthoryear{Gong \& Ji}{2018}]{gong}
Gong~Y-X., Ji~J., 2018, MNRAS, 478, 4565

\bibitem[\protect\citeauthoryear{Go{\'z}dziewski \oo}{2013}]{gozdzi}
Go{\'z}dziewski~K., S{\l}onina~M., Migaszewski~C.,~Rozenkiewicz~A.,
2013, MNRAS, 430, 533

\bibitem[\protect\citeauthoryear{Gray \& Brown}{2001}]{graybra}
Gray~D.~F., Brown~K., 2001, PASP, 113, 723

\bibitem[\protect\citeauthoryear{Griffin \& Redman}{1960}]{griffin}
Griffin~R.~F., Redman~R.~O., 1960, MNRAS, 120, 287

\bibitem[\protect\citeauthoryear{Hatzes, Cochran \& Bakker}{1998}]{hatzpeg}
Hatzes~A.~P., Cochran~W.~D., Bakker~E.~J., 1998, ApJ, 508, 380

\bibitem[\protect\citeauthoryear{Hatzes \oo}{2003}]{gamceph}
Hatzes~A.~P., Cochran~W.~D., Endl~M., McArthur~B., Paulson~D.~B., 
Walker~G.~A.~H., Campbell~B., Yang~S., 2003, ApJ, 599, 1383

\bibitem[\protect\citeauthoryear{Hatzes \oo}{2015}]{alde}
Hatzes~A.~P. \oo, 2015, A\&A, 580, A31

\bibitem[\protect\citeauthoryear{Hatzes \oo}{2018}]{gdra}
Hatzes~A.~P. \oo, 2018, AJ, 155, 120

\bibitem[\protect\citeauthoryear{Hearnshaw \oo}{2002}]{hercules}
Hearnshaw~J.~B., Barnes~S.~I., Kershaw~G.~M., Frost~N., Graham~G.,
Ritchie~R., Nankivell~G.~R., 2002, Exp. Astron., 13, 59

\bibitem[\protect\citeauthoryear{Horner \oo}{2019}]{horner}
Horner~J. \oo, 2019, AJ, 158, 100

\bibitem[\protect\citeauthoryear{Jefferys}{1974}]{jeff}
Jefferys~W.~H., 1974, AJ, 79, 710

\bibitem[\protect\astroncite{Kervella \oo}{2019}]{kervella}
Kervella~P., Arenou~F., Mignard~F., Th{\'e}venin~F., 2019, A\&A, 623, A72

\bibitem[\protect\astroncite{Kjeldsen \& Bedding}{1995}]{kjeldsen}
Kjeldsen~H., Bedding~T.R., 1995, A\&A, 293, 87

\bibitem[\protect\astroncite{Kovtyukh \oo}{2003}]{kov2003}
Kovtyukh~V.~V., Soubiran~C., Belik~S.~I., Gorlova~N.~I., 2003, A\&A, 411, 559

\bibitem[\protect\citeauthoryear{Kratter \& Perets}{2012}]{kratter}
Kratter~K.~M., Perets~H.~B., 2012, ApJ, 753, 91

\bibitem[\protect\citeauthoryear{K{\"u}rster \oo}{2003}]{barnard}
K{\"u}rster~M. \oo, 2003, A\&A, 403, 1077

\bibitem[\protect\citeauthoryear{LightKurve}{2018}]{lightkurve}
Lightkurve Collaboration \oo, 2018, Lightkurve: Kepler and TESS time series 
analysis in Python, Astrophysics Source Code Library, (ascl: 1218.013)

\bibitem[\protect\citeauthoryear{Lin, Bodenheimer \& Richardson}{1996}]{lin}
Lin~D.~N.~C., Bodenheimer~P., Richardson~D.~C., 1996, Nature, 380, 606

\bibitem[\protect\citeauthoryear{Linsky \& Avrett}{1970}]{linsky}
Linsky~J.~L., Avrett~E.~H., 1970, PASP, 82, 169

\bibitem[\protect\citeauthoryear{Marzari \& Thebault}{2019}]{galaxies}
Lisogorskyi~M., Jones~H.~R.~A., Feng~F., 2019, MNRAS, 485, 4804

\bibitem[\protect\citeauthoryear{Mayor \& Queloz}{1995}]{51pegA}
Mayor~M., Queloz~D.~A., 1995, Nature, 378, 355

\bibitem[\protect\citeauthoryear{Morais |& Guippone}{2012}]{morais}
Morais~M.~H.~M., Guippone~C.~A., 2012, MNRAS, 424, 52

\bibitem[\protect\citeauthoryear{Morton }{2000}]{morton}
Morton~D., 2000, ApJS, 130, 403

\bibitem[\protect\citeauthoryear{Narita \oo}{2009}]{hatP7b}
Narita~N., Sato~B., Hirano~T., Tamura~M., 2009, PASJ, 61, L35

\bibitem[\protect\citeauthoryear{Oranje}{1983}]{oranje}
Oranje~B.~J., 1983, A\&A, 124, 43

\bibitem[\protect\citeauthoryear{Ortiz \oo}{2016}]{ortiz}
Ortiz~M. \oo, 2016, A\&A, 595, A55

\bibitem[\protect\citeauthoryear{Panichi \oo}{2017}]{panichi}
Panichi~F., Go{\'z}dziewski~K., Turchetti~G., 2017, MNRAS, 468, 469

\bibitem[\protect\citeauthoryear{P{\'e}rez Mart{\'i}nez \oo}{2011}]{perez}
P{\'e}rez Mart{\'i}nez~M.~I., Schr\"{o}der~K.-P., Cuntz~M., 2011, MNRAS, 414,418

\bibitem[\protect\citeauthoryear{Press \oo}{1994}]{numrec}
Press~W.~H., Teukolsky~S.~A., Vetterling~W.~T., Flannery~B.~P., 1994, Numerical 
Recipes in C: The Art of Scientific Computing, Cambridge Uni.~Press, New York

\bibitem[\protect\citeauthoryear{Quarles \oo}{2012}]{quarlesa}
Quarles~B., Cuntz~M., Musielak~Z.~E., 2012, MNRAS, 421, 2930

\bibitem[\protect\citeauthoryear{Quarles \oo}{2020}]{quarlesb}
Quarles~B., Li~G., Kostov~V., Haghighipour~N., 2020, AJ, 159, 80

\bibitem[\protect\citeauthoryear{Ramm}{2004}]{thesis}
Ramm~D.~J., 2004, PhD thesis, Univ.~Canterbury

\bibitem[\protect\citeauthoryear{Ramm}{2015}]{djr15}
Ramm~D.~J., 2015, MNRAS, 449, 4428

\bibitem[\protect\citeauthoryear{Ramm \oo}{2009}]{djr09}
Ramm~D.~J., Pourbaix~D., Hearnshaw~J.~B., Komonjinda~S., 2009, MNRAS, 394, 
1695

\bibitem[\protect\citeauthoryear{Ramm \oo}{2016}]{djr16}
Ramm~D.~J. \oo, 2016, MNRAS, 460, 3706

\bibitem[\protect\citeauthoryear{Rasio \& Ford}{1996}]{rasio}
Rasio~F.~A., Ford~E.~B., 1996, Science, 274, 954

\bibitem[\protect\citeauthoryear{Reichert \oo}{2019}]{reichert}
Reichert~K., Reffert~S., Stock~S., Trifonov~T., Quirrenbach~A., 2019, A\&A, 
625, 22

\bibitem[\protect\citeauthoryear{Ricker \oo}{2015}]{ricker}
Ricker~G.~R. \oo, 2015, J. Astron. Telesc. Instrum. Syst., 1, 1

\bibitem[\protect\citeauthoryear{Robertson \oo}{2012}]{paul2012}
Robertson~P. \oo, 2012, ApJ, 749, 39

\bibitem[\protect\citeauthoryear{Robertson \oo}{2013}]{paul2013}
Robertson~P., Endl~M., Cochran~W.~D., Dodson-Robinson~S.~E., 2013, ApJ, 764, 3

\bibitem[\protect\citeauthoryear{Rutten}{1984}]{rutten}
Rutten~R.~G.~M., 1984, A\&A, 130, 353

\bibitem[\protect\citeauthoryear{Saio \oo}{2015}]{saio}
Saio~H., Wood~P.~R., Takayama~M., Ita~Y., 2015, MNRAS, 452, 3863 

\bibitem[\protect\citeauthoryear{Skuljan}{2004}]{hrsp}
Skuljan J., 2004, in Kurtz D. W., Pollard K., eds, Proc. IAU Coll. 193, 
Variable Stars in the Local Group, Astronomical Society of the Pacific, San 
Francisco, p.~575

\bibitem[\protect\citeauthoryear{Stock \oo}{2020}]{stock}
Stock~K., Cai~M.~X., Spurzem~R., Kouwenhoven~M.~B.~N., Zwart~S.~P., 2020, 
MNRAS, 497, 1807 

\bibitem[\protect\citeauthoryear{Sturrock \& Scargle}{2010}]{sturrock}
Sturrock~P.~A., Scargle~J.~D., 2010, ApJ, 718, 527

\bibitem[\protect\citeauthoryear{Trifonov \oo}{2018}]{hd59686}
Trifonov~T., Lee~M.~H., Reffert~S., Quirrenbach~A., 2018, AJ, 155, 257

\bibitem[\protect\citeauthoryear{Tutukov \& Ferorova}{2012}]{tutukov}
Tutukov~A.~V., Fedorova~A.~V., 2012, Astron.~Rep., 56, 305

\bibitem[\protect\citeauthoryear{Vanderburg \oo}{2020}]{vanderburg}
Vanderburg~A. \oo, 2020, Nature, 585, 363

\bibitem[\protect\citeauthoryear{Walker \oo}{1992}]{walker}
Walker~G.~A.~H., Bohlender~D.~A., Walker~A.~R., Irwin~A.~W., Yang~S.~L.~S., 
Larson~A., 1992, ApJ, 396, L91

\bibitem[\protect\citeauthoryear{Warner}{1969}]{warner}
Warner~B., 1969, MNRAS, 144, 333

\bibitem[\protect\citeauthoryear{Wiegert \& Holman}{1997}]{wiegert}
Wiegert~P.~A., Holman~M.~J., 1997, AJ, 113, 1445

\bibitem[\protect\citeauthoryear{Wilson}{1963}]{wilson63}
Wilson~O.~C., 1963, ApJ, 138, 832

\bibitem[\protect\citeauthoryear{Wilson}{1978}]{wilson78}
Wilson~O.~C., 1978, ApJ, 226, 379

\bibitem[\protect\citeauthoryear{Wilson \& Bappu}{1957}]{wilbap}
Wilson~O.~C., Bappu~M.~K.~V., 1957, ApJ., 125, 661

\bibitem[\protect\citeauthoryear{Winn \oo}{2009}]{winn}
Winn~J.~N., Johnson~J.~A., Albrecht~S., Howard~A.~W., Marcy~G.,W., 
Crossfield~I.~J., Holman~M.~J., 2009, ApJ, 703, L99

\bibitem[\protect\citeauthoryear{Wolszczan \& Frail}{1992}]{pulsar}
Wolszczan~A., Frail~D., 1992, Nature, 355, 145

\bibitem[\protect\citeauthoryear{Snowden}{1981}]{snowden}
Young~A., Furenlid~I., Snowden~M.~S., 1981, ApJ, 245, 998

\bibitem[\protect\citeauthoryear{Zechmeister \& K{\"u}rster}{2009}]{gls}
Zechmeister~M., K{\"u}rster~M., 2009, A\&A, 496, 577

\bibitem[\protect\citeauthoryear{Zechmeister \oo}{2018}]{zech}
Zechmeister~M. \oo, 2018, A\&A, 609,12

\end{thebibliography}
\end{document}